\documentclass[12pt,a4paper,english]{article}
\usepackage{titling}
\usepackage[affil-it]{authblk}
\usepackage{longtable}
\usepackage{lscape}
\usepackage{graphicx}
\usepackage{epsfig}
\usepackage{dsfont}
\usepackage{amsfonts}
\usepackage{amsmath}
\usepackage{amsthm}
\usepackage{amssymb}
\usepackage{bbold}
\usepackage[english]{babel}
\usepackage[utf8]{inputenc}
\usepackage[left=0.8in,right=0.8in,top=1.0in,bottom=1.2in]{geometry}
\usepackage{calc}
\usepackage{cancel}
\usepackage{float}
\usepackage{slashed}
\usepackage{mathrsfs}
\usepackage{pdfpages}
\usepackage{tabu}
\usepackage{simplewick}
\usepackage[hidelinks]{hyperref}
\usepackage{bm}
\usepackage{multirow}
\usepackage{hhline}
\usepackage{pifont}
\usepackage{float}
\usepackage[caption = false]{subfig}
\usepackage{xcolor}
\usepackage[normalem]{ulem}
\usepackage{cite}

\usepackage{datetime}
\usepackage{booktabs}

\numberwithin{equation}{section}
\numberwithin{figure}{section}
\numberwithin{table}{section}

\makeatletter
\g@addto@macro\bfseries{\boldmath}
\makeatother

\allowdisplaybreaks

\newcommand{\be}{\begin{equation}}
\newcommand{\ee}{\end{equation}}
\newcommand{\bea}{\begin{eqnarray}}
\newcommand{\eea}{\end{eqnarray}}
\newcommand{\ba}{\begin{align}}
\newcommand{\ea} {\end{align}}

\newcommand{\GF}{G_F}
\newcommand{\sw}{s_w^2}

\renewcommand{\O}{\mathcal{O}}
\renewcommand{\Re}{\text{Re}}

\newcommand{\xt}{x_t}
\newcommand{\xz}{x_Z}

\def\eq#1{{Eq.~(\ref{#1})}}
\def\eqs#1#2{{Eqs.~(\ref{#1})--(\ref{#2})}}
\def\fig#1{{Fig.~\ref{#1}}}

\def\Table#1{{Table~\ref{#1}}}

\def\sec#1{{Sect.~\ref{#1}}}

\def\app#1{{Appendix~\ref{#1}}}
\def\apps#1#2{{Apps.~\ref{#1}--\ref{#2}}}

\def\nn {\nonumber}

\DeclareMathOperator{\sgn}{sgn}

\allowdisplaybreaks

\newcommand{\published}[1]{%
\gdef\puB{#1}}
\newcommand{\puB}{}

\date{}
 
\begin{document}

\title{\textbf{Constraining flavour patterns of scalar leptoquarks in the effective field theory}}

\author[1,2]{Marzia Bordone\thanks{marzia.bordone@to.infn.it}}
\author[1]{Oscar Cat\`{a}\thanks{oscar.cata@uni-siegen.de}}
\author[1]{Thorsten Feldmann\thanks{thorsten.feldmann@uni-siegen.de}}
\author[1]{Rusa Mandal\thanks{rusa.mandal@uni-siegen.de}}
\affil[1]{Theoretische Physik 1, Naturwissenschaftlich-Technische Fakult\"at, Universit\"at Siegen, Walter-Flex-Stra{\ss}e 3, D-57068 Siegen, Germany}
\affil[2]{Dipartimento di Fisica, Universit\`a di Torino \& INFN, Sezione di Torino, I-10125 Torino, Italy}

\published{\flushright SI-HEP-2020-23, P3H-20-046\vskip2cm}

\maketitle

\begin{abstract}
We investigate the viability of extending the Standard Model with $S_1$ and $S_3$ scalar leptoquarks when the flavour structure is parametrized in terms of Froggatt-Nielsen charges. In contrast to a similar analysis with a vector leptoquark, we find essentially two solutions for the charges that fit the experimental constraints, which are dominated by the current tensions in $B$ decays. These two scenarios differ in their estimate of the anomalous magnetic moment of the muon, $(g-2)$, but they both predict sizeable contributions to $\tau\to\mu\gamma$, $\bar B_s\to\tau^\pm\mu^\mp$ and $B^+\to K^+\tau^+\mu^-$ decays, whose branching ratios are close to the current experimental limits. 
\end{abstract}

\clearpage
\tableofcontents
\clearpage

\section{Introduction}
\label{sec:1}
Effective field theories (EFTs) are one of the most efficient tools to explore physics at low energies.  Since their construction is general and independent of the details of physics at higher scales, they are extremely useful as discovery tools at high-energy colliders. In particular, EFTs at the electroweak scale provide a generic parametrization of new physics (NP) effects, and are used as a theoretical template for indirect searches of physics beyond the Standard Model (SM).

Depending on rather generic assumptions on the nature of NP and electroweak symmetry breaking, there exist two EFTs at the electroweak scale, the so-called Standard Model Effective Field
Theory (SMEFT)\cite{Buchmuller:1985jz,Grzadkowski:2010es} and the Electroweak Chiral Lagrangian (EW$\chi$L)~\cite{Feruglio:1992wf,Contino:2010mh,Alonso:2012px,Buchalla:2013rka}. Both EFTs have by now been developed at one-loop level and match the precision requirements of the LHC and its upgrades in the gauge and scalar sectors.

However, when flavour is added into the picture, things become more complicated. The main reason is that very little is known about the pattern of flavour symmetry breaking and, since each of the EFT coefficients is a tensor in flavour space, the number of unknown parameters becomes extremely large. A purely phenomenological approach of fitting the flavour pattern from experiment can only work when very few operators are involved, but it is clearly not viable for global flavour fits.

An alternative is to assume a particular source of flavour-symmetry breaking and introduce a flavour power-counting scheme using spurions. A paradigmatic example of this philosophy is minimal flavour violation (MFV)\cite{DAmbrosio:2002vsn,Buras:2003jf}. In this picture, the EFT consists of a double expansion, one related to gauge symmetries and one to flavour symmetries. In general, the choice of flavour spurions can be motivated by theoretical arguments, phenomenological hints, or a combination of both. MFV has the virtue of simplicity, since it considers the SM Yukawa matrices as the only flavour spurions, and the flavour power counting is inferred from the observed pattern of fermion masses and CKM angles. 

The procedure to generalize MFV is well-defined (see e.g. ref.~\cite{Feldmann:2006jk}), but a non-minimal flavour structure needs to be well-motivated and the issue of power counting has to be resolved. The discrepancies and tensions found in the last years in different observables in $B$ physics cannot be accommodated with MFV and therefore it is justified to explore non-minimal flavour scenarios.

In a previous paper~\cite{Bordone:2019uzc} we laid out the main ideas towards an EFT-based approach to NP with flavour. The key element of that analysis is a bottom-up flavour structure with a flavour power counting set by introducing Froggatt-Nielsen (FN) charges~\cite{Froggatt:1978nt} for the different fermion fields (for other studies that use FN charges in connection with leptoquark scenarios see~\cite{Varzielas:2015iva,Hiller:2016kry,deMedeirosVarzielas:2019lgb}). The flavour spurions are selected based on dynamical input. In~\cite{Bordone:2019uzc}, for instance, we singled out those associated with the exchange of a hypercharge $Y=\tfrac{2}{3}$ vector leptoquark $U_1$. With this selection of flavour spurions, the FN charges were constrained by phenomenological input from low-energy flavour observables. The resulting viable FN assignments described different phenomenological scenarios and, in particular, provided different predictions to be tested experimentally. 

This approach provides a simple and systematic way of introducing flavour structures into EFTs. Let us stress that this proposed framework is different from a pure EFT approach which, even with the FN power counting built in, would otherwise contain too many free parameters. That is why some top-down information from leptoquark models is kept. However, this top-down information is flavour-independent, and it is used merely to simplify our analysis. Instead, the  viable FN charges are determined entirely by low-energy data, i.e. they are not taken from an underlying theory of flavour, though an analysis of its regularities might lead to interesting theoretical connections. For recent applications of FN charges in different contexts of flavour physics, see e.g. refs.\cite{Smolkovic:2019jow,Fedele:2020fvh}. 

Besides the vector leptoquark $U_1$, different scalar leptoquark combinations can also successfully accommodate the current flavour tensions observed in both charged and neutral current decays of the $B$ meson. Among these, the combination of the two hypercharge $Y=\tfrac{1}{3}$ scalar leptoquarks, known as $S_1$ and $S_3$, has been singled out in the literature for its success in explaining low-energy data~\cite{Marzocca:2018wcf,Becirevic:2018afm,Bigaran:2019bqv,Crivellin:2019dwb,Saad:2020ihm,Gherardi:2020qhc,Babu:2020hun,Crivellin:2017zlb,Buttazzo:2017ixm}.
In this paper, we apply the procedure outlined in ref.~\cite{Bordone:2019uzc} to the flavour spurions associated with the scalar leptoquarks $S_1$ and $S_3$. We investigate the allowed patterns of FN charges through a fit to low-energy observables. In the case we are studying, high-energy observables, e.g. bounds from high-$p_T$ tails, are far less constraining~\cite{Greljo:2017vvb,Greljo:2018tzh,Marzocca:2020ueu}. An advantage of scalar leptoquarks (as compared to vector leptoquarks) is that they provide a renormalizable UV model, while scenarios with vector leptoquarks have to be embedded into e.g. some grand unified framework (for recent work in this direction see, for instance, refs.~\cite{Bordone:2017bld,Bordone:2018nbg,Cornella:2019hct,Blanke:2018sro,DiLuzio:2017vat,DiLuzio:2018zxy,Fornal:2018dqn}
). Observables which are loop-induced can therefore be reliably computed and also implemented in the fit. We stress that, even though working with scalars ensures the renormalisability of the theory, the flavour structure we use relies on the EFT description and FN power counting discussed in \cite{Bordone:2019uzc}.

We find that the FN charges for left-handed fermions are basically fixed by the experimental constraints, with values that essentially match those of the $U_1$ setting. As expected, deviations occur for the FN charges of the right-handed fermions, since the $U_1$ leptoquark and the $S_1$ leptoquark couple to the right-handed down-quark and up-quark sectors, respectively. We find a small number of phenomenologically allowed solutions for the scalar leptoquarks, with very similar qualitative features. This is in  contrast to the $U_1$ case, where different solutions lead to rather distinct phenomenological scenarios~\cite{Bordone:2019uzc}.

The most relevant prediction of our setting is a large value for $\tau\to\mu\gamma$, basically at the edge of the current experimental limits. Also $\bar B_s\to\tau^\pm\mu^\mp$ and $B^+\to K^+\tau^+\mu^-$ modes are enhanced and they are predicted to have a branching ratio one order of magnitude smaller than the present bounds. Lepton flavour-conserving modes such as $\bar B_s\to\tau^+\tau^-$ and $B^+\to K^+\tau^+\tau^-$ also show one order of magnitude enhancement compared to their SM expectation. However, the current experimental limits are quite weak. On the other hand, we find a rather modest correction to the muon $(g-2)$, typically ten times smaller than the observed deviation between the experimental measurement and the SM prediction. While this can -- in principle --
be fixed by tuning some of the overall coupling coefficients and lowering the value for the leptoquarks masses, given the then arising tension with $\tau\to\mu\gamma$, it does not seem to be a natural outcome of our setup. 

This paper is organized as follows: in Sect.~\ref{sec:2} we introduce the scalar leptoquarks and summarize their relevant interactions. In Sect.~\ref{sec:3} we discuss the power counting based on FN charges and the constraints to make it compatible with the SM. The list of low-energy flavour observables to be considered is discussed in Sect.~\ref{sec:4}, together with the allowed textures for the flavour spurions. A fit is performed in Sect.~\ref{sec:5}, where we comment on the predictions of our setting and possible further experimental tests. Conclusions are given in Sect.~\ref{sec:6}. Technical details are compiled in two Appendices.       

\section{Flavour spurions for scalar leptoquarks}
\label{sec:2}

Flavour spurions can be conveniently classified according to their representation under the maximal flavour symmetry of the SM that commutes with the gauge symmetries, namely\footnote{Additional $U(1)$ factors are not shown for simplicity.}
\begin{align}
{\cal{G}}_f=SU(3)_Q\times SU(3)_U\times SU(3)_D\times SU(3)_L\times SU(3)_E\,.
\end{align}
If one considers the fermionic content of the theory to be SM-like, then there are 44 different flavour spurions associated with Dirac bilinears~\cite{Bordone:2019uzc}. Since we are interested in leptoquark scenarios, the list gets reduced to 12 spurions.

It is natural to understand these different spurions as originating from the interactions of heavy leptoquarks and SM fermions. Each leptoquark model then has a definite number of associated spurions. In ref.~\cite{Bordone:2019uzc} we studied the phenomenology of the hypercharge $Y=\tfrac{2}{3}$ vector leptoquark $U_1$, and consequently examined the spurions $\Delta_{QL}$ and $\Delta_{DE}$. In this work we are interested in a scenario with scalar leptoquarks. Previous studies~\cite{Marzocca:2018wcf,Becirevic:2018afm,Bigaran:2019bqv,Crivellin:2019dwb,Saad:2020ihm,Gherardi:2020qhc,Babu:2020hun,Crivellin:2017zlb,Buttazzo:2017ixm} have shown that the tensions observed in $B$ physics can be accommodated with the two (weak singlet and triplet) hypercharge $Y=\tfrac{1}{3}$ scalar leptoquarks. These scalars are typically referred to in the literature as $S_1$ and $S_3$.    

Once $S_1$ and $S_3$ are included, the SM Lagrangian is enlarged to \cite{Dorsner:2016wpm}
\begin{equation}
\begin{aligned}
\mathcal{L}&=\mathcal{L}_{\rm{SM}}+D_\mu S^{a\dagger}_3 D^\mu S^{a}_3-M_3^2S^{a\dagger}_3 S^a_3+D_\mu S_1^\dagger D^\mu S_1-M_1^2S_1^\dagger S_1-V(S_3,S_1,H)\\
&+g_3 \tilde{S}_{QL}^{i\alpha}\bar{Q}^{ci}\epsilon \sigma^a L^\alpha  S_3^a+ g_1 S^{i\alpha}_{QL}  \bar{Q}^{ci}\, \epsilon L^\alpha\, S_1 + g_R S^{i\alpha}_{UE} \;  \bar{ u}^{ci}_R \, \ell_R^\alpha \; S_1\; + \mathrm{h.c.} \,,
\label{eq:S1S3L}
\end{aligned}
\end{equation}
where $\epsilon = i\sigma_2$ is the antisymmetric isospin tensor, $Q$ and $L$ are the left-handed quark and lepton doublets, respectively, and $Q^c = i\gamma_0\gamma_2\bar{Q}^T$ denotes 
the charge-conjugated Dirac fields. The covariant derivatives are defined as
\begin{equation}
\begin{aligned}
D_\mu S_1=&\left(\partial_\mu-ig_s G_\mu^AT^A -i\frac{g'}{3}B_\mu\right)S_1\,, \\
D_\mu S_3=&\left(\partial_\mu-ig_s G_\mu^AT^A -ig W_\mu^k I^k -i\frac{g'}{3}B_\mu\right)S_3\,,
\end{aligned}
\end{equation}
where $T^A$ and $I^k$ are the $SU(3)_C$ and $SU(2)_L$ generators, respectively.
In \eq{eq:S1S3L} the indices for the flavour matrices $S_{QL}$, $\tilde{S}_{QL}$ and $S_{UE}$ are shown explicitly. The term $V(S_3,S_1,H)$ encodes the interacting potential between the Higgs boson and the leptoquarks. It is relevant in studies of the electroweak vacuum stability~\cite{Bandyopadhyay:2016oif}, oblique corrections~\cite{Gherardi:2020det} or Higgs boson decays~\cite{Crivellin:2020tsz}. A study of this term is beyond the scope of this work and therefore we ignore it for simplicity.

Both $S_1$ and $S_3$ are in principle allowed to couple to quark bilinears which violate baryon number and in particular induce proton decay. In this paper we enforce baryon number conservation at the TeV scale. Baryon number violating operators are therefore set to vanish.

We choose to work in the down-quark basis for the $SU(2)_L$ doublets, namely
\begin{equation}
Q^i \equiv 
\begin{pmatrix}
V^{*}_{ij} u^j \\
d^i
\end{pmatrix}\,,
\qquad
L^\alpha \equiv 
\begin{pmatrix}
\nu^\alpha \\
e^\alpha
\end{pmatrix}\,,
\label{eq:down_alignment}
\end{equation}
where $V$ denotes the CKM mixing matrix. Working with the charge eigenstates for the leptoquarks, defined through $S_3^1 = (S_3^{4/3}+S_3^{-2/3})/\sqrt{2}$, $S_3^2 = i(S_3^{4/3}-S_3^{-2/3})/\sqrt{2}$ and $S_3^3\equiv S_3^{1/3}$, the interaction Lagrangian thus reads
\begin{align}
\mathcal{L}_\text{int}^{F} &=\, g_3  \bigg[\sqrt{2} (V^* \tilde S_{QL})^{i\alpha} \bar{u}_L^{ci}\nu_L^\alpha S_3^{-2/3}- \sqrt{2}  \tilde S_{QL}^{i\alpha}\bar{d}_L^{ci}e_L^\alpha S_3^{4/3}-\tilde S_{QL}^{i\alpha}\bar{d}_L^{ci}\nu_L^\alpha S_3^{1/3}-(V^* \tilde S_{QL})^{i\alpha}\bar{u}_L^{ci}e_L^\alpha S_3^{1/3}\bigg]\nonumber\\
&+g_1\left[ (V^* S_{QL})^{i\alpha} \,  {\bar{u}^{ci}_{L\!}}  e_L^\alpha - S_{QL}^{i\alpha} \, {\bar{d}^{ci}_{L\!}}\,  \nu_L^\alpha\right]S_1+g_R  S_{UE}^{i\alpha}\, {\overline{ u}^{ci}_{R\!}}\,  e_R^\alpha  S_1 + \text{h.c.}\,,\\
\mathcal{L}_\text{int}^{G}&=-i g (W^+_\mu {\cal{J}}_-^\mu+W^-_\mu {\cal{J}}_+^\mu)+ieA_\mu {\cal{J}}^\mu_A+i \frac{g}{\cos \theta_{W}}Z_\mu {\cal{J}}^\mu_Z\,,
\end{align}
with
\begin{align}
{\cal{J}}_+^\mu&=S_3^{4/3}\stackrel{\leftrightarrow}{\partial_\mu} S_{3}^{1/3*}+S_3^{1/3}\stackrel{\leftrightarrow}{\partial_\mu} S_{3}^{-2/3*}\,,\\
{\cal{J}}_-^\mu&=S_3^{-2/3}\stackrel{\leftrightarrow}{\partial_\mu} S_{3}^{1/3*}+S_3^{1/3}\stackrel{\leftrightarrow}{\partial_\mu} S_{3}^{4/3*}\,,\\
{\cal{J}}_A^\mu&=\sum_j Q_jS_3^{j}\stackrel{\leftrightarrow}{\partial_\mu} S_{3}^{j*}+\frac{1}{3}S_1\stackrel{\leftrightarrow}{\partial_\mu} S_1^{*}\,,\\
{\cal{J}}_Z^\mu&=\sum_j(t_3^j-Q_j\sin^2\theta_W)S_3^{j}\stackrel{\leftrightarrow}{\partial_\mu} S_{3}^{j*}-\frac{1}{3}\sin^2\theta_W S_1\stackrel{\leftrightarrow}{\partial_\mu} S_1^{*}\,,
\end{align}
where $A\!\stackrel{\leftrightarrow}{\partial_\mu}\!B=A\,\partial_\mu B-B\,\partial_\mu A$ and we have omitted the gluonic interactions, which are relevant for collider observables at high-$p_T$ but do not play a key role in our analysis.

In order to accommodate the tensions in $B$ physics, the scalar leptoquarks have to be relatively light, around the TeV scale. At the electroweak scale they can therefore be integrated out and matched onto the SMEFT. The tree-level matching is shown in \eq{eq:lagrangian_SMEFT}, where the canonical basis of ref.~\cite{Grzadkowski:2010es} is used. The matching coefficients read
\begin{align}
[\mathcal{C}_{lq}^{(1)}]^{ij\alpha\beta} &= -\frac{1}{4}(3\, |g_3|^2 \tilde S_{QL}^{j\beta} \tilde S_{QL}^{*i\alpha} +|g_1|^2 S_{QL}^{j\beta} S_{QL}^{*i\alpha} )\,, \\
[\mathcal{C}_{lq}^{(3)}]^{ij\alpha\beta} &= -\frac{1}{4}( |g_3|^2 \tilde S_{QL}^{j\beta}\tilde S_{QL}^{*i\alpha} -|g_1|^2 S_{QL}^{j\beta} S_{QL}^{*i\alpha})\,, \\
[\mathcal{C}_{eu}]^{ij\alpha\beta} &= -\frac{1}{2}  |g_R|^2S_{UE}^{j\beta} S_{UE}^{*i\alpha} \,, \\
[\mathcal{C}_{lequ}^{(1)}]^{ij\alpha\beta} &= -4 \, [\mathcal{C}_{lequ}^{(3)}]^{ij\alpha\beta} = \frac{1}{2} g_R\, g_1^* S_{QL}^{*j\beta} S_{UE}^{i\alpha} \,.
\label{eq:matching_2LQ}
\end{align}

\section{EFT approach with flavour power-counting}
\label{sec:3}

A valid power counting has two basic requirements: (i)  it reproduces the SM flavour structure with a simple setup, and (ii) the scheme is self-consistent, in particular the addition of spurions and their combinations does not upset the hierarchies already present in the SM. This second requirement is highly nontrivial and in practice strongly constrains the form of the power-counting scheme. 

In ref.~\cite{Bordone:2019uzc} we adopted a power counting based on the well-known Froggatt and Nielsen (FN) model~\cite{Froggatt:1978nt}. The FN model is a theory of flavour that introduces a (spontaneously broken) new $U(1)$ symmetry with generation-dependent $U(1)$ charge assignments to each quark multiplet. Provided that a sufficient number of heavy fermions exists and that spontaneous breaking is triggered by the vacuum expectation value $\langle \phi_{\rm FN}\rangle $ of a new scalar field at a scale $\Lambda_{\rm FN} \gg \langle \phi_{\rm FN}\rangle$, 
the model can accommodate the SM flavour hierarchies. Flavour non-diagonal transitions are suppressed by powers of $\lambda=(\langle \phi_{\rm FN}\rangle/\Lambda_{\rm FN}) \ll 1$, which is usually associated with the Cabibbo angle, $\lambda \approx \sin^2\theta_{C} \approx 0.2$, and their magnitude is determined by the corresponding FN charges. The suggestion made in ref.~\cite{Bordone:2019uzc} (see also ref.~\cite{Feldmann:2006jk}) is to merely assign FN charges to the SM fields, without addressing the problem of which dynamical mechanism can generate them. With this prescription it is then rather straightforward to take also leptons into account.

With the generalised FN prescription, every flavour structure in the SMEFT is determined only by the difference of FN charges of the fields present. For instance, if we denote the fermion FN charges by $b_Q^i, b_D^i, b_U^i$ and $b_L^\alpha, b_E^\alpha$ in a flavour basis defined by the $U(1)$ symmetry of the FN construction (FN basis), the SMEFT operator
\begin{align}
[\mathcal{C}_{lequ}^{(1)}]^{ij\alpha\beta}(\bar{Q}^i u_R^j)\epsilon(\bar{L}^\alpha e^\beta_R)
\end{align}
has the flavour scaling
\begin{align}
[\mathcal{C}_{lequ}^{(1)}]^{ij\alpha\beta}\sim \lambda^{|b_Q^i-b_L^\alpha|+|b_U^j-b_E^\beta|}\,,
\end{align}
where the form of this factorized structure keeps track of the fact that the FN mechanism is linked to a spurion decomposition of fermion bilinears.

Before we move to the NP contributions, the flavour structure of the SM (fermion masses and CKM mixing angles) already sets constraints on some combinations of the FN charges. Concerning the latter, the CKM matrix can be written, in terms of FN charges, as
\begin{align}
V_{ij}=(V_{U_L}^{\dagger}V_{D_L})_{ij} 
\sim \lambda^{\vert b_Q^i-b_Q^j\vert} \,, 
\end{align}
where $V_X$ denote the rotation matrices from the flavour to the mass eigenbasis for a given quark species. The left-handed quark charges are fixed by matching the previous expression onto the generally accepted Wolfenstein parametrization of the CKM matrix:
\begin{equation}\label{CKM}
V\sim \left(
\begin{array}{ccc}
1 & \lambda & \lambda^3\\
\lambda & 1 & \lambda^2\\
\lambda^3 & \lambda^2 & 1
\end{array}
\right)\,.
\end{equation} 
Since only the absolute value of charge differences can be constrained, the FN charges $b_Q^i$ are fixed up to a common offset $d$ and an absolute sign. The general solutions
\begin{align}
b_Q=(3+d,2+d,d)\,\qquad\textrm{and}\qquad b_Q=(3+d,4+d,6+d)
\end{align}
simply expose this inherent ambiguity. The two solutions above actually differ by a global sign flip of all charges only. The parameter $d$ sets the values of the $U(1)_{FN}$ charges in a would-be FN model. In the phenomenological approach we are using in this paper, this absolute value is of no significance. For simplicity, we choose 
\begin{align}
 b_Q^1 \equiv 3 \,, \qquad b_Q^2 \equiv 2 \,, \qquad b_Q^3 \equiv 0 \,.
\end{align}
In turn, the entries of the Yukawa matrices scale as 
\begin{align}
(Y_U)_{ij}\sim& \lambda^{\vert b_Q^i-b_U^j\vert}\,, \qquad  
(Y_D)_{ij}\sim \lambda^{\vert b_Q^i-b_U^j\vert}\,,\qquad  
(Y_E)_{\alpha\beta}\sim \lambda^{\vert b_L^\alpha-b_E^\beta \vert}\,.
\end{align}
One of the features of a power counting based on FN charges is their basis-independence. We can therefore set bounds on the charges by working in the fermion mass eigenbasis.

Concerning the eigenvalues of the quark Yukawa matrices, we have 
\begin{align}
 y_u & \sim \lambda^{|b_Q^1-b_U^1|} \approx \lambda^8 \,, &y_d & \sim \lambda^{|b_Q^1-b_D^1|} \approx \lambda^7 \,,\cr 
 y_c& \sim \lambda^{|b_Q^2-b_U^2|} \approx \lambda^4 \,, & y_s& \sim \lambda^{|b_Q^2-b_D^2|} \approx \lambda^5 \,, \cr 
 y_t & \sim \lambda^{|b_Q^3-b_U^3|} \approx \lambda^0 \,, &y_b & \sim \lambda^{|b_Q^3-b_D^3|} \approx \lambda^3 \,.
\end{align}
The above expressions fix the right-handed quark FN charges up to a twofold ambiguity:
\begin{equation}
\begin{array}{lll}
b_U^1 \simeq-5,+11  \,, & \quad
b_U^2 \simeq-2,+6 \,, & \quad
b_U^3 \equiv 0  \,,\cr\cr
b_D^1 \simeq-4,+10  \,, & \quad
b_D^2 \simeq-3,+7 \,, & \quad
b_D^3 \equiv -3,+3  \,.	
\end{array}
\end{equation}

The leptonic FN charges are considerably less constrained by the SM. The masses of the charged leptons require
\begin{align}
\label{eq:lepton_SM_FN}
 y_e & \sim \lambda^{|b_L^1-b_E^1|} \approx \lambda^9\,,\quad
 &y_\mu& \sim \lambda^{|b_L^2-b_E^2|} \approx \lambda^5\,,\quad
 &y_\tau & \sim \lambda^{|b_L^3-b_E^3|} \approx \lambda^3 \,.
\end{align}
Additional flavour spurions are sensitive to different combinations of FN charges and lead to further constraints. The phenomenologically allowed values for the FN charges therefore depend on both the phenomenological input and the spurions used to accommodate it. 
In the present paper we enlarge the SM flavour structure with the following
spurions, defined as:
\begin{align}
\tilde S_{QL}^{i\alpha}= \tilde c_L^{i\alpha} \,  \lambda^{|b_Q^i-b_L^\alpha|}  \label{eq:spurion1}\,, \\
S_{QL}^{i\alpha}= c_L^{i\alpha} \,  \lambda^{|b_Q^i-b_L^\alpha|} \label{eq:spurion2}\,,\\
S_{UE}^{i\alpha}= c_R^{i\alpha} \, \lambda^{|b_U^i-b_E^\alpha|}  \label{eq:spurion3}\,.
\end{align}
The extra constraints on the FN charges from low-energy phenomenology for these specific spurions are discussed in the next section.
\section{Methodology and observables}
\label{sec:4}
Even with a well-defined FN power counting, the number of free parameters describing the NP spurions is substantial. Beside the FN charges, we have the flavour-dependent coefficients $c_L^{i\alpha}$, $\tilde{c}_L^{i\alpha}$ and $c_R^{i\alpha}$. By construction, these coefficients are assumed to be ${\cal{O}}(1)$ complex numbers. The power counting limits their magnitude but does not reduce the number of them. 

Notice that we can consider the FN charges as integers without loss of generality: the effects of non-integer charges can be completely absorbed into a redefinition of the coefficients $c_j^{i\alpha}$. Likewise, a shift of $\lambda$ can also be absorbed by $c_j^{i\alpha}$. In order to simplify our analysis, we make the following assumptions:
\begin{itemize}
\item The masses of the leptoquarks are assumed to be degenerate and we set their value (the cutoff scale of the EFT) to be $M= 2~\text{TeV}$. With this conservative choice we avoid the constraints from direct searches, e.g. from high-$p_T$ tails, which set a lower limit on the leptoquarks masses of $\sim 1~$TeV  (see e.g. refs.~\cite{Greljo:2017vvb,Greljo:2018tzh,Marzocca:2020ueu} for a detailed EFT-based analysis and ref.~\cite{Angelescu:2018tyl,Bandyopadhyay:2018syt,Mandal:2018kau} for specific leptoquark benchmarks).

\item The spurion entries are assumed to be real (in the FN basis) and flavour-universal up to a relative sign, which is dictated by phenomenological requirements. In practice, we set $c_j^{i\alpha}=\pm 1$ and the only free coefficients left are $g_1$, $g_3$ and $g_R$ defined in \eq{eq:S1S3L}, assumed to be real. In particular, this implies that any source of CP violation comes from the SM parameters only. The flavour-independence of the coefficients is a rather strong requirement, motivated only to reduce the number of free parameters in our analysis. In any realistic theory these  coefficients are  flavour-dependent. In the following sections we  discuss in which cases this requirement has to be relaxed.

\item The CKM matrix elements are taken from the NP fit by the UTFit collaboration~\cite{UTFit}. In the fit that we perform in \sec{sec:5} we fix the CKM parameters to their central values, without treating them as nuisance parameters. We expect the size of the error associated with this simplified procedure to be negligible.

\end{itemize}
With the previous assumptions, the free parameters are the (integer) FN charges, the relative signs of the spurion entries, and the overall coupling strengths $g_j$ in the Wilson coefficients. The latter are expected to be $\O(1)$ numbers. 

From the previous discussion, it is clear that in our approach there is nothing fundamental about the values of the FN charges, since  their values are correlated with the choices for $M$, $\lambda$ and the simplifying assumptions for the Wilson coefficients. Any interpretation of their values should therefore be taken with a grain of salt.

All of the previous assumptions can be gradually lifted, once more and more precise data become available. In this work we take a very simplified setting, which is justified by the existence of a number of viable solutions (to be discussed below). Relaxing the assumptions above would definitely increase the number of allowed scenarios, but this would be hardly informative, except to indicate that the current data is not precise enough to discriminate among the scenarios. 

\subsection{Low-energy observables}

Scalar leptoquarks contribute to a rich set of low-energy processes. As a consequence of our power counting, we are sensitive to observables which involve fermions belonging to any of the three families. Since the Lagrangian in \eq{eq:S1S3L} is renormalisable, processes induced at one-loop order are also  considered in our analysis. In the following we list the most stringent constraints on the spurion entries. In the next subsection we translate them into constraints on the FN charges. In order to facilitate this comparison, all the dimensionless parameters are expressed as integer powers of $\lambda$.

We first discuss the modes that show deviations from the SM, and therefore justify the introduction of the leptoquark flavour spurions. We then comment on the most relevant modes that set limiting constraints.
  
\begin{enumerate}
\item[(i)]{$\bm{R_{D^{(*)}}}$:} The measured values of the lepton-flavour universality ratios $R_{D^{(*)}}$ currently show a tension of $3-4\, \sigma$ with respect to their SM predictions (references to experimental measurements and SM predictions are given in \app{app:B1}). They are therefore one of the most sizeable effects that leptoquark scenarios have to accommodate. Using the matching in \eq{eq:matching_2LQ} and the expression in \eq{eq:observable: RDst}, we have
\begin{equation}
\begin{aligned}
R_{D^{(*)}} \approx R_{D^{(*)}}|_\text{SM}\bigg[1-&\frac{v^2}{2 M^2}\sum_{j=1}^3 \frac{V_{cj}}{V_{cb}}(|g_3|^2 \tilde S_{QL}^{33}\tilde S_{QL}^{*j3}-|g_1|^2 S_{QL}^{33}S_{QL}^{*j3}) \\
+&\frac{v^2}{2 M^2}\sum_{j=1}^3 \frac{V_{cj}}{V_{cb}}(|g_3|^2 \tilde S_{QL}^{32}\tilde S_{QL}^{*j2}-|g_1|^2 S_{QL}^{32}S_{QL}^{*j2})\\
-&\frac{v^2}{4 M^2 V_{cb}}g_1g_R^*\left(\mathcal{F}^{VS}_{D^{(*)}}(\tau)-\frac{1}{4}\mathcal{F}^{VT}_{D^{(*)}}(\tau)\right)S_{QL}^{33}S_{UE}^{*23}\bigg]\,.
\end{aligned}
\label{eq:RD}
\end{equation}
In order to be consistent with $b\to c\ell\bar{\nu}$ data, with $\ell = \mu,e$, we assume that the correction is mostly driven by the couplings of the third generation (first and third lines in \eq{eq:RD}). A correction of ${\cal{O}}(10\%)$ with respect to the SM translates into 
\begin{align}
\left|\sum_{i=1}^3 \frac{V_{ib}}{V_{cb}}(|g_3|^2 \tilde S_{QL}^{33}\tilde S_{QL}^{*i3}-|g_1|^2 S_{QL}^{33}S_{QL}^{*i3}) \right|&\sim {\cal{O}}(\lambda^{-1}) \,,\\
\left|g_1g_R^*\left(\mathcal{F}^{VS}_{D^{(*)}}(\tau)-\frac{1}{4}\mathcal{F}^{VT}_{D^{(*)}}(\tau)\right)S_{QL}^{33}S_{UE}^{*23}\right|&\sim {\cal{O}}(\lambda) \,.
\end{align}
Expanding the first equation, one finds that 
\begin{align}
S_{QL}^{33}\left[S_{QL}^{*23}-\lambda S_{QL}^{*13}+ \lambda^2 S_{QL}^{*33}\right]\sim {\cal{O}}(\lambda) \,,  
\end{align}
which shows that the term proportional to $S_{QL}^{*23}$ is the dominant one. The same holds for the spurion $\tilde S_{QL}$. The constraints on the spurion matrices therefore read
\begin{equation}
\tilde S_{QL}^{33} \tilde S_{QL}^{*23} \sim \lambda\,,\quad S_{QL}^{33} S_{QL}^{*23} \sim \lambda\,, \quad S_{QL}^{33} S_{UE}^{*23} \sim \lambda \,.
\label{eq:gen_bounds_RD}
\end{equation}
In order to make sure that the contributions of both leptoquarks do not cancel each other, we  require 
\begin{equation}
 \tilde{S}_{QL}^{33} \tilde{S}_{QL}^{*23}< 0\qquad  \text{and } \qquad S_{QL}^{33} S_{QL}^{*23}>0\,.
\end{equation}

\item[(ii)]{$\bm{b\to s\ell^+\ell^-}$:} Global fits to $b\to s\mu^+\mu^-$ show that the most favoured scenario is given by $\mathcal{C}_9^{2322} = -\mathcal{C}_{10}^{2322} \approx -0.5$, where the coefficients are defined in \eq{eq:base_NC}, assuming that $b\to se^+e^-$ is SM-like~\cite{Alguero:2019ptt,Aebischer:2019mlg,Ciuchini:2019usw,Hurth:2020rzx}. At the tree level, only $S_3$ contributes. Using \eq{eq:observables_bsll} for the tree-level matching, one finds
\begin{equation}
\tilde S_{QL}^{*32}\tilde S_{QL}^{22}\sim \lambda^4\,,
\label{eq:bounds_gen_bsmumu}
\end{equation} 
for the muon mode and 
\begin{equation}
\tilde S_{QL}^{*31}\tilde S_{QL}^{21}< \lambda^6\,,
\end{equation}
for the electron mode.
 
The contribution of $S_1$ is one-loop suppressed and proceeds via box diagrams which are negligible in our framework. A similar suppression applies to the one-loop contributions from $S_3$, which we likewise neglect.

We have checked that the current bounds on $\bar{B}_s\to\tau^+\tau^-$ decay are less constraining and therefore do not add additional information to our analysis.
\\
In ref.\cite{Crivellin:2018yvo} it has been shown that a universal shift in $\mathcal{C}_9^{23ii}$, with $i=1,\,2,\,3$, can be obtained through penguin-type diagrams with a $\tau$ lepton in the loop. In our framework these contributions are small and hence we can neglect them. Accordingly, $\mathcal{C}_9^{2322}$ contains only tree-level NP effects.

\item[(iii)]{$\bm{Z\to \nu \bar\nu$} and $\bm{Z\to \ell^+\ell^-$}}: LEP bounds provide a test of NP effects in $Z$ couplings to leptons. In our setup, such corrections arise from penguin diagrams. Compared to the results of global fits (recently updated in ref.~\cite{Falkowski:2019hvp}) we find that in our setting scalar-leptoquark effects are generically rather suppressed, with the top contribution being the dominant one. The strongest constraint comes from $Z\to\nu\bar{\nu}$, which is bound by the measurement on the effective number of neutrino species $N_\nu^\text{exp}$. We find
\begin{equation}
N_\nu^\text{exp}= \sum_{ij} \left|\delta_{ij} + \frac{\delta g_{\nu_L}^{ij}}{g_\nu^\text{SM}}\right\vert^2 = 2.9963 \pm 0.0074 \,,
\end{equation}
where the expression for $\delta g_{\nu_L}^{ij}$ is given in Eq.~\eqref{eq:znunu_expression}. Note that the contributions to $N_\nu^\text{exp}$ from the $S_3$ and $S_1$ leptoquarks are always positive, which is a generic result, independent of considerations on flavour power counting. This excess can be acceptable as long as
\begin{align}
|\tilde S_{QL}^{3\alpha}|^2\geq \lambda^2\,.
\end{align}
This is not the case for $Z\to \ell^+\ell^-$ decays, where the direction of the NP corrections matches the experimental trend. Uncertainties are however relatively large and the resulting constraints on the spurion entries are less stringent. An interesting exception is $Z\to \tau^+_L \tau^-_L$, where one finds the following conditions:
\begin{align}
|\tilde{S}_{QL}^{33}|^2\geq \lambda^2\,, \qquad|S_{QL}^{33}|^2\geq \lambda^2\,.
\end{align}

\item[(iv)]{$\bm{K^+\to\pi^+\nu\bar\nu}$:} 
Using the expressions in \eq{eq:WCKtopinunu} and the limit from Eq.~\eqref{eq:RKpinunu}, we have
\begin{eqnarray}
\sum_\alpha  \frac{v^2}{M^2}\frac{\pi \sin^2\theta_W}{\alpha_\text{EM}}\left\vert \frac{1}{y_\nu}\left(|g_3|^2 \tilde S_{QL}^{1\alpha} \tilde S_{QL}^{*2\alpha}+|g_1|^2 S_{QL}^{1\alpha} S_{QL}^{*2\alpha}\right)\right\vert< 0.69\,, \label{eq:Ktopinunu_limit1}\\
\sum_{\alpha \ne \beta}  \frac{v^2}{M^2}\frac{\pi \sin^2\theta_W}{\alpha_\text{EM}}\left\vert \frac{1}{y_\nu}\left(|g_3|^2 \tilde S_{QL}^{1\alpha} \tilde S_{QL}^{*2\beta}+|g_1|^2 S_{QL}^{1\alpha} S_{QL}^{*2\beta}\right)\right\vert< 0.46\,,\label{eq:Ktopinunu_limit2}
\end{eqnarray}
where \eq{eq:Ktopinunu_limit1} corresponds to the case where neutrinos have the same flavour and \eq{eq:Ktopinunu_limit2} encodes the LFV contributions.
With $y_\nu\sim \lambda^3$, one finds the following bounds on the spurions:
\begin{equation}
 \tilde{S}_{QL}^{1\alpha} \tilde{S}_{QL}^{*2\beta}<\lambda^5\quad \text{and} \quad 
 S_{QL}^{1\alpha} S_{QL}^{*2\beta}<\lambda^5\,.
 \label{eq:gen_bounds_Ktopi}
\end{equation}

\item[(v)]{$\bm{\bar{B}\to K^{(*)}\nu\bar\nu}$:} using the limits in \eq{eq:limitsBKnunu} and the expressions in \eqs{eq:matching_BKnunu}{eq:BtoKnunu}, we have
\begin{eqnarray}
\sum_\alpha \frac{v^2}{M^2} \frac{\pi}{\alpha_\text{EM}|V_{tb} V_{ts}^*|} \left\vert\frac{1}{C_{BK}^\text{SM}}\left(|g_3|^2 \tilde S_{QL}^{2\alpha} \tilde S_{QL}^{*3\alpha}+|g_1|^2 S_{QL}^{2\alpha} S_{QL}^{*3\alpha}\right)\right\vert<5.1\,, \label{eq:BKnunu_bound1} \\
\sum_{\alpha\neq \beta} \frac{v^2}{M^2} \frac{\pi}{\alpha_\text{EM}|V_{tb} V_{ts}^*|} \left\vert\frac{1}{C_{BK}^\text{SM}}\left(|g_3|^2 \tilde S_{QL}^{2\alpha} \tilde S_{QL}^{*3\beta}+|g_1|^2 S_{QL}^{2\alpha} S_{QL}^{*3\beta}\right)\right\vert<2.6\,,\label{eq:BKnunu_bound2} 
\end{eqnarray}
where \eq{eq:BKnunu_bound1} and \eq{eq:BKnunu_bound2} correspond to the flavour-conserving and flavour-violating contributions, respectively. With $C_{BK}^\text{SM}\approx -6.35$, the previous conditions imply the constraints:
\begin{equation}
\tilde{S}_{QL}^{2\alpha} \tilde{S}_{QL}^{*3\beta}<\lambda^2 \quad \text{and} \quad  S_{QL}^{2\alpha} S_{QL}^{*3\beta}<\lambda^2 \,,
\label{eq:bounds_gen_BtoK}
\end{equation}
assuming that the terms from $S_3$ and $S_1$ do not cancel each other. Depending on the structure of $\tilde{S}_{QL}^{i\alpha}$ and $S_{QL}^{i\alpha}$, the LFV contributions can be numerically important. This  turns out to be the case for some of the allowed solutions to be discussed later on. In this case, the LFV contributions have to be included in the numerical analysis.

\item[(vi)]{$\bm{\Delta M_q}$}: neutral meson mixings are very sensitive probes of NP effects. In particular, in $B_d-\bar{B}_d$ and $B_s-\bar{B}_s$ mixing the leading contributions are proportional to the spurion combinations (see \app{app:Bmixing})
\begin{align}
(\tilde{S}_{QL}^{*j3} \tilde{S}_{QL}^{33})^2 \,, \quad (S_{QL}^{*j3} S_{QL}^{33})^2\,, \quad (\tilde{S}_{QL}^{*j3} \tilde{S}_{QL}^{33})(S_{QL}^{*j3} S_{QL}^{33})\,,
\end{align}
where the three terms correspond to the different ways that $S_1$ and $S_3$ can be exchanged in box diagrams and $j=1,2$ correspond to the different light quarks.
The bounds that we obtain are
\begin{align}
\tilde{S}_{QL}^{* j3} \tilde{S}_{QL}^{33}\geq \lambda^3\,, \qquad S_{QL}^{*j3} S_{QL}^{33}\geq \lambda^3\,.
\end{align}
We notice that for $\Delta M_s$ these bounds are not fulfilled, since the same spurion combination appears in the description of $R_{D^{(*)}}$. Therefore, we require the term with mixed spurions to have a negative interference with the others in order to partially reduce the tension. In contrast, the spurion combination relevant for $\Delta M_d$ is not present in any other of the considered observables, and one needs a more global analysis to understand whether the constraint can be fulfilled or partial cancellations are required.   
 
The contributions with pure muon and pure electron exchange in the loop, which are constrained by $b\to s\mu^+\mu^-$ and $b\to s e^+ e^-$, are within the experimental bounds for $\Delta M_s$. The effect of the diagrams with different lepton species in the loop turns out to be numerically suppressed due to the bounds from other LFV decays.

\item[(vii)]{$\bm{K_L\to\mu e }$}: The bound on this decay mode provides a stringent constraint on the spurion entries associated with light lepton generations. Comparing the current experimental upper bound with the expression in Eq.~\eqref{eq:Ptoll}, one requires that 
\begin{equation}
\tilde S_{QL}^{*21}\tilde S_{QL}^{12}\leq \lambda^8\, \qquad \text{and} \qquad \tilde S_{QL}^{*22}\tilde S_{QL}^{11}\leq \lambda^8\,.
\end{equation}

\item[(viii)]{$\bm{\mu\to e\gamma}$:} The current upper bound also constrains significantly the spurion entries for light leptons. Comparing the expression in \eq{eq:muegamma} with the current experimental bound we find, from the left-handed operators,
\begin{equation}
\begin{aligned}
|\lambda^3 \tilde{S}_{QL}^{12}+\lambda^2 \tilde{S}_{QL}^{22}+\tilde{S}_{QL}^{32}| |\lambda^3 \tilde{S}_{QL}^{*11}+\lambda^2 \tilde{S}_{QL}^{*21}+\tilde{S}_{QL}^{*31}|\leqslant \lambda^4 \,,\\
|\lambda^3 S_{QL}^{12}+\lambda^2 S_{QL}^{22}+S_{QL}^{32}| |\lambda^3 S_{QL}^{*11}+\lambda^2 S_{QL}^{*21}+S_{QL}^{*31}|\leqslant \lambda^4 \,, 
\end{aligned}
\end{equation}
which implies the minimal conditions:
\begin{equation}
\tilde{S}_{QL}^{32} \tilde{S}_{QL}^{*31}  \leqslant \lambda^4 \,, \qquad S_{QL}^{32} S_{QL}^{*31}  \leqslant \lambda^4 \,.
\end{equation}
More stringent limits come from the scalar and tensor contributions induced by $S_1$, due to the chiral enhancement of the quark loops. From the top contribution we extract the conservative bounds:
\begin{equation}
\label{eq:mu2e_LR}
S_{QL}^{32} S_{UE}^{*31}  \leqslant \lambda^{10} \,,\qquad S_{QL}^{*31} S_{UE}^{32}  \leqslant \lambda^{10}\,.
\end{equation}

\item[(ix)]{$\bm{\tau\to \mu\gamma}$:} The leading contribution comes from the scalar and tensor operators, just as in $\mu\to e\gamma$. However, the $\tau$ upper limits are experimentally less constrained, and one finds the conditions
\begin{equation}
S_{QL}^{33} S_{UE}^{*32}  \leqslant \lambda^{4} \,,\qquad S_{QL}^{*32} S_{UE}^{33}  \leqslant \lambda^{4}\,.
\end{equation}

\end{enumerate}

From the previous analysis of the different processes, one immediately identifies some tensions that are generic to the $S_1$+$S_3$ leptoquarks scenario. The corrections to $b\to s\ell^+\ell^-$ can in general be implemented without upsetting other processes. The main problem comes with $R_{D^{(*)}}$. The combinations $\tilde{S}_{QL}^{33}\tilde{S}_{QL}^{*23}$, $S_{QL}^{33} S_{QL}^{*23}$ and $S_{QL}^{33}S_{UE}^{*23}$ have to be sizeable to match the experimental measurements. The first two combinations generate large corrections to neutral meson mixings, $K\to\pi\nu\bar{\nu}$ and $B\to K^{(*)}\nu\bar{\nu}$ decays, imposing a necessary partial cancellations between the $S_1$ and $S_3$ leptoquark contributions. The last combination leads to large effects on the LFV decays $\mu\to e\gamma$ and $\tau\to\mu\gamma$. Since this combination is only generated by $S_1$, a suppression cannot rely on partial cancellations and it is typically harder to achieve. Some other tensions affect $Z\to\nu{\bar{\nu}}$ or $W$ lepton-flavour universality tests (discussed in \apps{app:Ztoll}{app:WLFU}). These tensions are however different in nature: the scalar leptoquarks generate NP contributions to these modes with a definite sign, which in some cases happens to conflict with the one from the global fits to experimental data.

In the following subsection we investigate the implications of these tensions for the FN power counting.  

\subsection{Constraints on the FN charges} 
\label{sec:FN_constr}

The conditions on FN charges discussed in \sec{sec:3} are linked to the SM flavour structure. This alone determines the charges of the quark fields (up to twofold ambiguities for the right-handed fields). The FN charges can be further constrained by using the low-energy observables listed above. This procedure clearly depends on the experimental situation and the selection of spurions. The additional constraints that we derive below are therefore associated with the specific extension chosen in this paper and the current experimental situation. The increase or decrease of certain tensions and improvement on certain bounds would in general lead to different values of the charges. An important point to stress is that compliance with flavour tests does not single out a solution for the FN charges, i.e.\ there is some leftover freedom and a range of values are allowed. This leaves us with a manageable number of potential solutions, which we analyse in the fit discussed in \sec{sec:5}.

Processes sensitive to $S_{QL}S_{QL}$ and $\tilde{S}_{QL}\tilde{S}_{QL}$ can be used to set constraints on the left-handed lepton charges, while processes that get contributions from scalar and tensor operators (proportional to $S_{QL}S_{UE}$) are useful to constrain the right-handed charges. Contributions proportional to $S_{UE}S_{UE}$ affect up-quark sector processes. Since these are rather weakly constrained, one finds no extra condition on the FN charges.

\begin{enumerate}
\item {\bf{$b_L$ charges}}.

\begin{itemize}
\item[(a)] Constraints on all families come from the bounds on $Z\to \nu{\bar{\nu}}$. This requires that $\left(\delta g^Z_{\nu_L}\right)_{\alpha\alpha}\leq \lambda^2$, which translates into $|b_Q^3-b_L^\alpha| \geq 1$. Using that $b_Q^3=0$, we get the condition 
\begin{align}
  \left| b_L^\alpha \right| \geq 1\,.
\label{eq:FN_const1}
\end{align}
Further generic constraints come also from $K^+\to\pi^+\nu\bar\nu$ and $B\to K^{(*)}\nu\bar\nu$. \eqs{eq:gen_bounds_Ktopi}{eq:bounds_gen_BtoK} are tantamount to $|b_Q^1 -b_L^\alpha|+|b_Q^2-b_L^\alpha|\geq 5$ and $|b_Q^2-b_L^\alpha|+|b_Q^3-b_L^\alpha|\geq\ 2$. Both constraints can be fulfilled with
\begin{align}
b_L^\alpha\leq 0 \,\quad {\mathrm{or}} \,\quad b_L^\alpha\geq 5\,.
\label{eq:FN_const2} 
\end{align}
Overall the allowed solutions are $b_L^\alpha\leq -1$ and $b_L^\alpha\geq 5$. 

\item[(b)] Constraints specific to $b_L^2$ come from $b\to s\mu^+\mu^-$. From \eq{eq:bounds_gen_bsmumu} we find $|b_Q^2-b_L^2|+|b_Q^3-b_L^2|\sim\, 4$, which, using the values of the $b_Q$ charges, leads to
\begin{equation}
b_L^2 = -1,+3\,.
\end{equation}
Combined with the constraints in (a) above, only the solution $b_L^2=-1$ is viable.

\item[(c)] Constraints specific to $b_L^1$ come from the LFV modes $K_L\to \mu^{\pm} e^{\mp}$ and $\mu\to e\gamma$. The bounds for $K_L\to \mu^{\pm} e^{\mp}$ discussed in the previous section can be written as the conditions $|b_Q^1-b_L^1|+|b_Q^2-b_L^2|>8$ and $|b_Q^2-b_L^1|+|b_Q^1-b_L^2|>8$. Using the values for $b_Q^1$, $b_Q^2$ and $b_L^2=-1$ discussed above, this implies the range 
\begin{align}
b_L^1\leq -2 \,\quad {\mathrm{or}} \,\quad b_L^1\geq 8 \,.
\end{align}
The condition from $\mu\to e\gamma$ reads $|b_Q^3-b_L^2|+|b_Q^3-b_L^1| \geq 5$, which leads to
\begin{equation}
\left| b_L^1 \right| \geq 4\,.
\end{equation}
The allowed values are therefore $b_L^1\leq -4$ and $b_L^1\geq 8$. We have checked that the bound on $b\to s e^+e^-$ is automatically fulfilled and thus brings no further constraints.

\item[(d)] Constraints specific to $b_L^3$ come from $b\to c\tau^-\bar\nu$. From \eq{eq:gen_bounds_RD} one finds the relation $|b_Q^3-b_L^3|+|b_Q^2-b_L^3|\sim 1$. However, with the values for $b_Q^i$, it is easy to see that the previous constraint has no solution for (integer) $b_L^3$. We therefore set $|b_Q^3-b_L^3|+|b_Q^2-b_L^3|\sim 2$, from which one concludes that
\begin{align}
b_L^3 = 0,+1,+2\,.
\end{align}
\end{itemize}

Notice that the previous values for $b_L^3$ are incompatible with the generic bounds from \eqs{eq:FN_const1}{eq:FN_const2}. The easiest solution is to select $b_L^3=1$, which respects the bounds from $Z\to \nu{\bar{\nu}}$ and $B\to K^{(*)}\nu\bar\nu$, and then bring $K^+\to\pi^+\nu\bar\nu$  into the allowed experimental bounds by imposing a cancellation between the $S_1$ and $S_3$ contributions. This indicates that both $S_1$ and $S_3$ leptoquarks are needed in order to obtain a satisfactory description of low-energy data. The choice $b_L^3=2$ requires a much stronger fine-tuning and we dismiss it.  

The first condition on the coefficients' signs is therefore
\begin{equation}
\sgn(\tilde c_L^{1\alpha} \tilde c_L^{2\alpha})=-\sgn(c_L^{1\alpha} c_L^{2\alpha})\,.
\end{equation}
There is also a generic tension between $R_{D^{(*)}}$ and $B_s-\bar{B}_s$ mixing, as already mentioned in the previous subsection. Actually, with the allowed FN charges, the NP contribution would induce a correction of typically a few $\%$. This tension can be resolved by enforcing 
\begin{align}
\sgn(\tilde{c}_L^{23} \tilde{c}_L^{33}  c_L^{23} c_L^{33})=-1\,.
\end{align} 
This condition contains the same spurion combination that appears in $R_{D^{(*)}}$ and is actually automatically fulfilled if  
\begin{equation}
(\tilde{c}_L^{23} \tilde{c}_L^{33})<0,\qquad (c_L^{23} c_L^{33})>0 \,,
\end{equation}
which ensures that the contributions of $S_3$ and $S_1$ to $R_{D^{(*)}}$ have a positive interference with the SM, as data indicates.

When trying to fit all the observables (see \sec{sec:5}) it turns out that the contribution to $B_d$ mixing can be more easily accommodated if there is also a partial cancellation. The condition in this case reads
\begin{align}
\sgn(\tilde{c}_L^{13} \tilde{c}_L^{33}  c_L^{13} c_L^{33})=-1 \,.    
\end{align}
Finally, since the contributions to $B\to K^{(*)}\nu\bar\nu$ are very close to the experimental bounds, we additionally impose
\begin{equation}
\sgn(\tilde c_L^{2\alpha} \tilde c_L^{3\beta})=-\sgn(c_L^{2\alpha} c_L^{3\beta})\,.
\end{equation}

Considering all the above discussion, the FN charges for left-handed leptons are constrained to the values
\begin{equation}
\label{eq:bLcharges}
\begin{array}{lll}
b_L^1\leq -4 \, \lor  \, b_L^1\geq 8\,, &\qquad b_L^2=-1\,, &\qquad b_L^3=+1\,.
\end{array}
\end{equation}
Therefore, all the FN charges for the left-handed fermion fields are fixed, with the exception of $b_L^1$. It is interesting to remark that the FN charges that we found happen to be essentially the same as the ones obtained for a $U_1$ leptoquark~\cite{Bordone:2019uzc}. This is not entirely surprising, given that the flavour spurions $\Delta_{QL}$, $S_{QL}$ and $\tilde S_{QL}$ couple to the same fermion fields and therefore have the same combinations of FN charges. However, since in ref.~\cite{Bordone:2019uzc} $\mu\to e\gamma$, $B\to K^{(*)}\nu\bar\nu$ and $K^+\to\pi^+\nu\bar\nu$ did not play a significant role, the impact of the phenomenological input used in both cases is rather different, in particular for the bounds on $b_L^1$.  

\item {\bf{$b_U$ and $b_E$ charges}}: The choices for the $b_L$ charges found above can be combined with the SM conditions in \eq{eq:lepton_SM_FN} to narrow down the parameter space for the $b_E$ charges. One finds
\begin{align}
b_E^1=b_L^1\pm 9\,,\quad b_E^2=-6,4\,, \quad b_E^3=-2,4\,.
\end{align}
The contributions of scalar operators in the different processes, i.e., those contributions proportional to $S_{QL}S_{UE}$, provide additional constraints on both $b_U$ and $b_E$. The most stringent ones come from $R_{D^{(*)}}$, $\mu\to e\gamma$ and $\tau\to \mu\gamma$. 

A sizeable scalar contribution to $R_{D^{(*)}}$ sets the constraint $|b_U^2-b_E^3|\leq 1$, which can be fulfilled only with the combination  
\begin{equation}
b_U^2=-2\,,\qquad b_E^3=-2\,,
\end{equation}
together with the sign constraint
\begin{align}
c_L^{33}c_R^{23}<0\,.
\end{align}
The conditions listed in \eq{eq:mu2e_LR} for $\mu\to e\gamma$ become $|b_L^2-b_Q^3|+|b_E^1-b_U^3|\geq 10$ and $|b_L^1-b_Q^3|+|b_E^2-b_U^3|\geq 10$. The first one sets
\begin{align}
\left|b_E^1\right|\geq 9\,,
\end{align} 
while the second one is trivially satisfied with large enough $b_L^1$. In order to have conservative bounds, and in compliance with \eq{eq:bLcharges}, we choose $b_L^1=-5$ and $b_L^1=+8$ as our benchmark solutions.

In turn, the constraints for $\tau\to \mu\gamma$ become $|b_Q^3-b_L^3|+|b_U^3-b_E^2|\geq 4$ and $|b_Q^3-b_L^2|+|b_U^3-b_E^3|\geq 4$. The first one is trivially satisfied, while the second one cannot be fulfilled with $b_E^3=-2$. This tension between $R_{D^{(*)}}$ and $\tau\to \mu\gamma$ can only be resolved by relaxing our assumption that the Wilson coefficients associated to the spurions are flavour-independent (see the discussion below).

\end{enumerate}

In summary, with both SM and NP constraints taken into account, we find the FN charges for the quark sector
\begin{equation}
\begin{aligned}
b_Q^1&=3\,, & b_Q^2&=2\,, & b_Q^3&=0\,, \\
b_U^1&=-5,11\,, & b_U^2&=-2\,, & b_U^3&=0\,.
\end{aligned}
\end{equation}
For the lepton sector, the constraints read
\begin{equation}
\begin{aligned}
&b_L^2=-1\,, \qquad\qquad\quad b_L^3=1\,, \qquad\qquad\quad b_E^3=-2\,,\\
&(b_L^1;b_E^1,b_E^2)=\left\{(-5;-14,-6),\ (8;17,-6),\ (8;17,4)\right\}\,.
\end{aligned}
\end{equation}
In comparison with the $U_1$ case studied in ref.~\cite{Bordone:2019uzc}, where 11 scenarios were possible, the scenario with scalar leptoquarks is more constraining and only 6 solutions are possible. In both cases the left-handed charges are essentially fixed, up to $b_L^1$. The main difference comes from the fermionic right-handed sector. Since the scalar leptoquarks are sensitive to the up-quark sector, they are more tightly constrained by $R_{D^{(*)}}$. For the same reason, the phenomenological impact affects mostly charm physics, where uncertainties are large. As a result, and as is discussed in the next Section, there are little phenomenological differences between the 6 potential solutions.    
The main difference arises from the choice of $b_E^2$.
With this in mind, we identify the following two (non-degenerate) benchmark scenarios:
\begin{align}
&{\text{Scenario A}}: \qquad (b_L^1;b_U^1;b_E^1,b_E^2)=(-5;+11;-14,-6)\,,\\
&{\text{Scenario B}}: \qquad (b_L^1;b_U^1;b_E^1,b_E^2)=(+8;-5;+17,+4)\,.
\label{eq:charges_final}
\end{align}
In both scenarios we have chosen the value of $b_U^1$ that gives the strongest suppression to processes involving the first generation. We have checked explicitly that the phenomenology of the remaining possibilities is qualitatively very similar.

Regarding the sign constraints listed above, they do not lead to a unique solution. A minimal choice is to take the entries $\tilde c_L^{33}$, $c_L^{13}$, $c_L^{32}$ and $c_R^{23}$ negative. The form of the spurion matrices for the different scenarios therefore reads: 
\begin{equation}
\begin{aligned}
    \tilde S_{QL}^{(A)}&\sim
    \begin{pmatrix}
    \lambda^8 & \lambda^4 & \lambda^2 \\
    \lambda^7 & \lambda^3 & \lambda \\
    \lambda^5 & \lambda & -\lambda \\
    \end{pmatrix}\,,
     & 
     \tilde S_{QL}^{(B)}&\sim
    \begin{pmatrix}
    \lambda^5 & \lambda^4 & \lambda^2 \\
    \lambda^6 & \lambda^3 & \lambda \\
    \lambda^8 & \lambda & -\lambda \\
    \end{pmatrix}\,,
    \\
    S_{QL}^{(A)}&\sim
    \begin{pmatrix}
    \lambda^8 & \lambda^4 & -\lambda^2 \\
    \lambda^7 & \lambda^3 & \lambda \\
    \lambda^5 & -\lambda & \lambda \\
    \end{pmatrix}\,,
    & 
    S_{QL}^{(B)}&\sim 
    \begin{pmatrix}
    \lambda^5 & \lambda^4 & -\lambda^2 \\
    \lambda^6 & \lambda^3 & \lambda \\
    \lambda^8 & -\lambda & \lambda \\
    \end{pmatrix}\,,
    \\
     S^{(A)}_{UE} &\sim 
    \begin{pmatrix}
    \lambda^{25} & \lambda^{17} & \lambda^{13} \\
    \lambda^{12} & \lambda^4 & -1 \\
    \lambda^{14} & \lambda^6 & \lambda^2 \\
    \end{pmatrix}\,,
    & 
    S^{(B)}_{UE} &\sim 
    \begin{pmatrix}
    \lambda^{22} & \lambda^{9} & \lambda^3 \\
    \lambda^{19} & \lambda^6 & -1 \\
    \lambda^{17} & \lambda^4 & \lambda^2 \\
    \end{pmatrix}\,.
    \end{aligned}
\end{equation}

In order to study the viability and specific phenomenological features of these solutions, in the next Section we perform a fit. Note that the previous matrices still contain the assumption that the Wilson coefficients are flavour-independent, i.e. $|c^{i\alpha}|=1$. As we show in the next Section, this assumption has to be relaxed for the entries $c_L^{32}$ and $c_R^{33}$, otherwise the tension between $R_{D^{(*)}}$ and $\tau\to\mu\gamma$ cannot be resolved.

\section{Results and discussion}
\label{sec:5}
The viability of our framework is tested by performing a fit to a set of observables. Our fit procedure employs the probabilistic programming package \texttt{PyMC3} \cite{pymc3}, which uses the principles of Bayesian inference. The main ingredient for the analysis is the likelihood function $\mathcal{L}$, defined as 
\begin{equation}
\log(\mathcal{L}) = -\frac{1}{2}\sum_{i \, \in \,\text{obs} } \left(\mathcal{O}^i_{th}-\mathcal{O}^i_{exp}\right)^T \Sigma_i^{-1} \left(\mathcal{O}^i_{th}-\mathcal{O}^i_{exp}\right) \,,
\end{equation}
where  $\Sigma_i$ is the combination of theoretical and experimental covariance matrices for each observable $\mathcal{O}_i$. 

The results of our analysis are obtained in terms of posteria distributions for the parameters $g_1$, $g_3$ and $g_R$.
We include in our fit all observables for which there is currently a measurement. This includes $R_{D^{(*)}}$, $F_L^{D^*}$, universality in $b\to c\ell\bar\nu$ ($\ell =\mu,e$), the inputs from $b\to s\ell^+\ell^-$ global fits, $\Delta M_{d,s}$, $K^+\to\pi^+\nu\nu$, the effective number of light neutrino species $N_\nu$, and the corrections to $Z$ and $W$ couplings to leptons. Agreement with observables for which there exist only upper bounds is checked a posteriori.

In the following we discuss two fit benchmarks: Benchmark I has the spurions $S_{QL}$ and $\tilde{S}_{QL}$ only and accordingly the couplings $g_1$ and $g_3$ as free parameters. Benchmark II includes also the spurion $S_{UE}$ and the free parameters are $g_1$, $g_3$ and $g_R$. In both cases we choose Scenario A as our nominal fit. The differences with Scenario B are discussed later on in the text.

\subsection{Benchmark I}
We first investigate the possibility of setting $g_R = 0$. In this case Scenarios A and B are indistinguishable. 
We note that the likelihood for this benchmark is insensitive to  
the sign of  $g_{1}$ and $g_{3}$. For convenience, we restrict ourselves to the case where $g_{1,3}$ are both positive, but analogous results are obtained for the other three sign combinations.

\begin{figure}[!ht]
\centering
\includegraphics[scale=0.5]{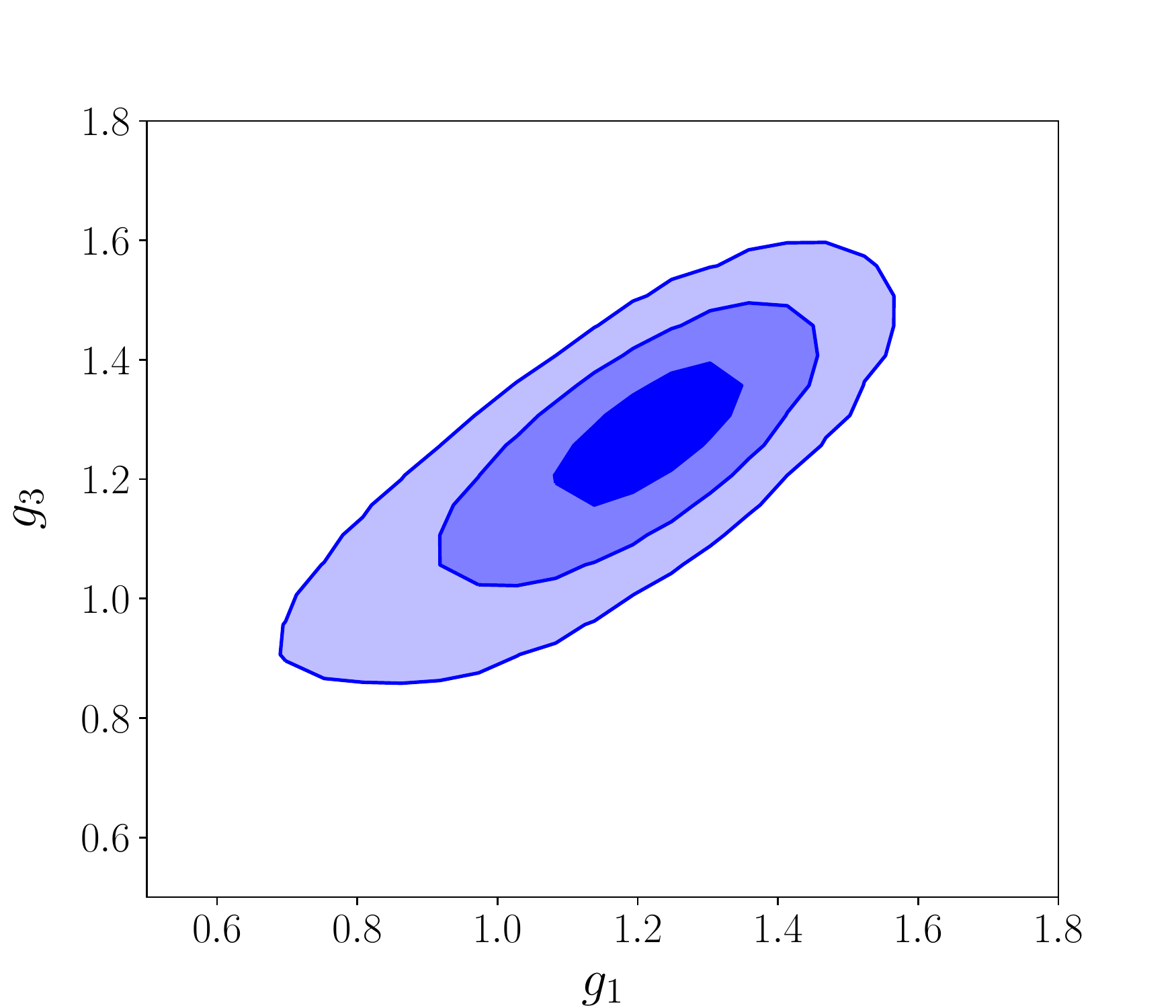}
\label{fig:chi2LH}
\caption{Fit results for the parameters $g_1$ and $g_3$ in Benchmark I ($g_R=0$). The shaded areas, from darker to lighter blue, describe the $1$, $2$ and $3 \, \sigma$ regions, respectively.} 
\end{figure}

Imposing flat priors for $g_1$ and $g_3$ we obtain the following mode and $68\%$ interval for our fit  parameters:
\begin{equation}
\begin{aligned}
g_1:\quad\text{mode}  = 1.25 \quad 68\%\,\, \text{interval}:[1.07,1.31]\,, \\
g_3:\quad\text{mode}  = 1.31 \quad 68\%\,\, \text{interval}:[1.15,1.36]\,,
\end{aligned}    
\end{equation}
with a correlation coefficient of $\rho_{g_1g_3} = 0.92$ (see \fig{fig:chi2LH}). Since the spurions $S_{QL}$ and $\tilde{S}_{QL}$ have the same power-counting scaling, the large correlation between $g_1$ and $g_3$ clearly indicates that observables with a significant impact on the likelihood scale as the combination $|g_1|^2+|g_3^2|$. Observables which scale like $|g_1|^2-|g_3^2|$ are instead very suppressed, and are mostly SM-like.

To evaluate the improvement of our model with respect to the SM hypothesis, we estimate the $\chi^2$, which is defined as $\chi^2 =-2 \log(\mathcal{L})$. In the SM hypothesis, the biggest contributions to $\chi^2 _\text{SM}$ come from $b\to s\ell^+\ell^-$ and $b\to c\tau\bar\nu$ data. The minimum of the $\chi^2$ corresponds to the point at which the parameters $g_{1,3}$ are exactly the estimated modes. In the minimum, we find $\Delta\chi^2 = \chi^2 _\text{SM}-\chi^2_\text{min}\sim 30$. 

The posteria are used to evaluate the NP contribution to various observables. In particular, we obtain that
\begin{equation}
R_{D^{(*)}} \sim 1.02\,R_{D^{(*)}}^\text{SM}  \,,
\end{equation}
which is too small to have a sizeable reduction of the observed tension in these observables. This shows the need to include scalar and tensor operators.

Before moving to Benchmark II, we comment on the corrections to the $W$ couplings $\delta g_W^{ij}$ defined in \eq{eq:Wcorrections}. Experimentally, one finds that $\delta g_W^{\alpha\alpha}$, with $\alpha=1,2,3$ shows some tension with the SM prediction. In particular, the highest discrepancy comes from $\delta g_W^{33}$, which is positive, while $\delta g_W^{11}$ and $\delta g_W^{22}$ are negative. The three modes combined give a contribution of $\sim 17$ to $\chi^2_\text{SM}$~\cite{Falkowski:2019hvp}.

In our framework, we find that the contributions from the $S_1$ and the $S_3$ leptoquarks (\eqs{eq:Wloop_S1}{eq:Wloop_S3}, respectively) largely cancel each other. The remaining overall correction is negative. However, it is so suppressed that the contribution to the $\chi^2_\text{min}$ from these modes is practically equivalent to the SM one.

We want to stress that this is not a specific feature of our framework: different scenarios, even with different mediators, share this issue. If one uses instead the recent ATLAS measurement of $R(\tau/\mu) = \mathcal{B}(W
\to\tau\bar\nu)/\mathcal{B}(W\to\mu\bar\nu)$ ~\cite{Aad:2020ayz}, which turns out to be much closer to the SM prediction, the contribution to the $\chi^2$ is very low, about $\sim 0.4$. New measurements of the $W$ couplings are thus definitely needed to understand if these tensions, observed mainly at LEP, are still significant or not. The present discussion also holds for Benchmark II, since the observables under consideration are independent of right-handed couplings.

\subsection{Benchmark II}
\begin{figure}[!ht]
\centering
\includegraphics[scale=0.45]{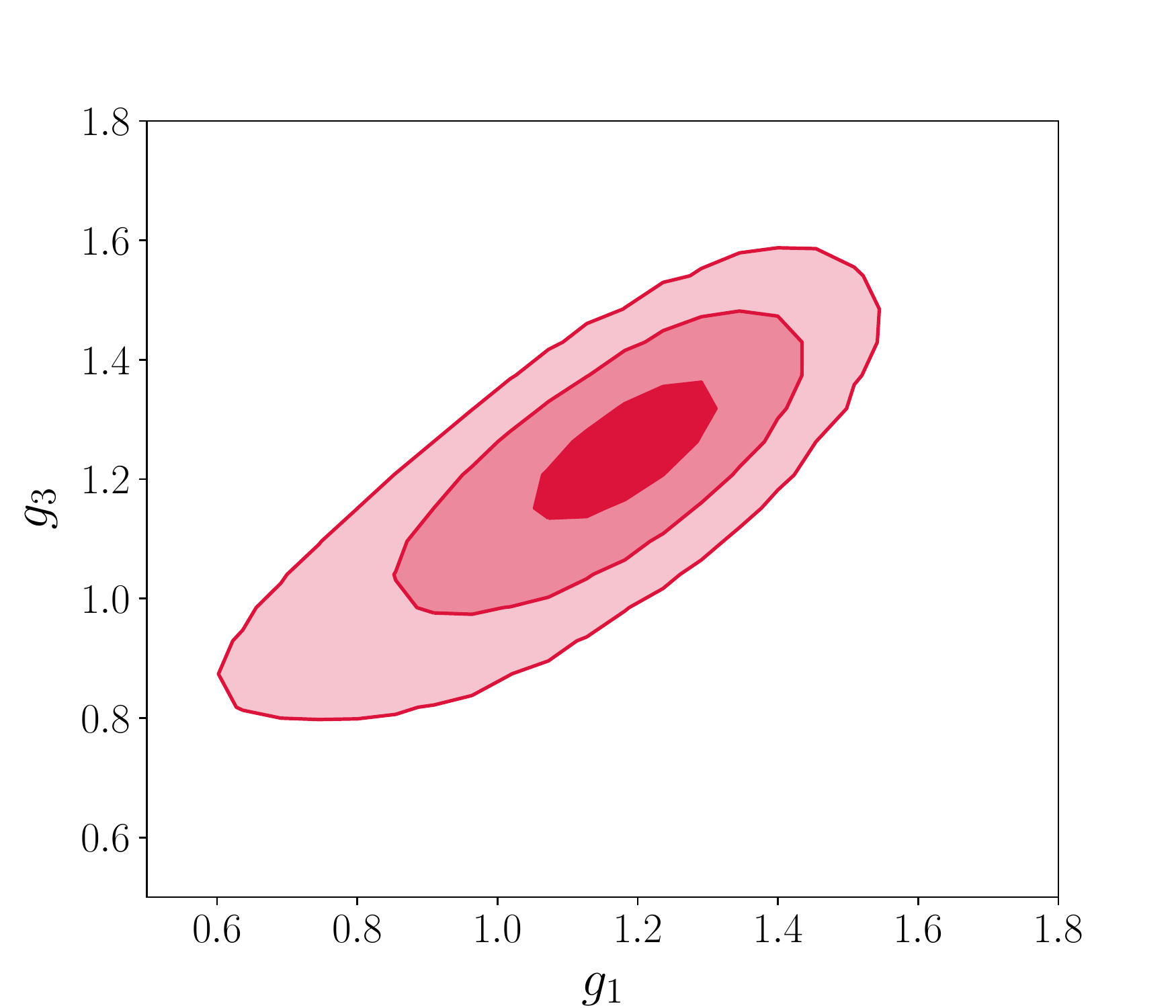} \hskip 5pt
\includegraphics[scale=0.45]{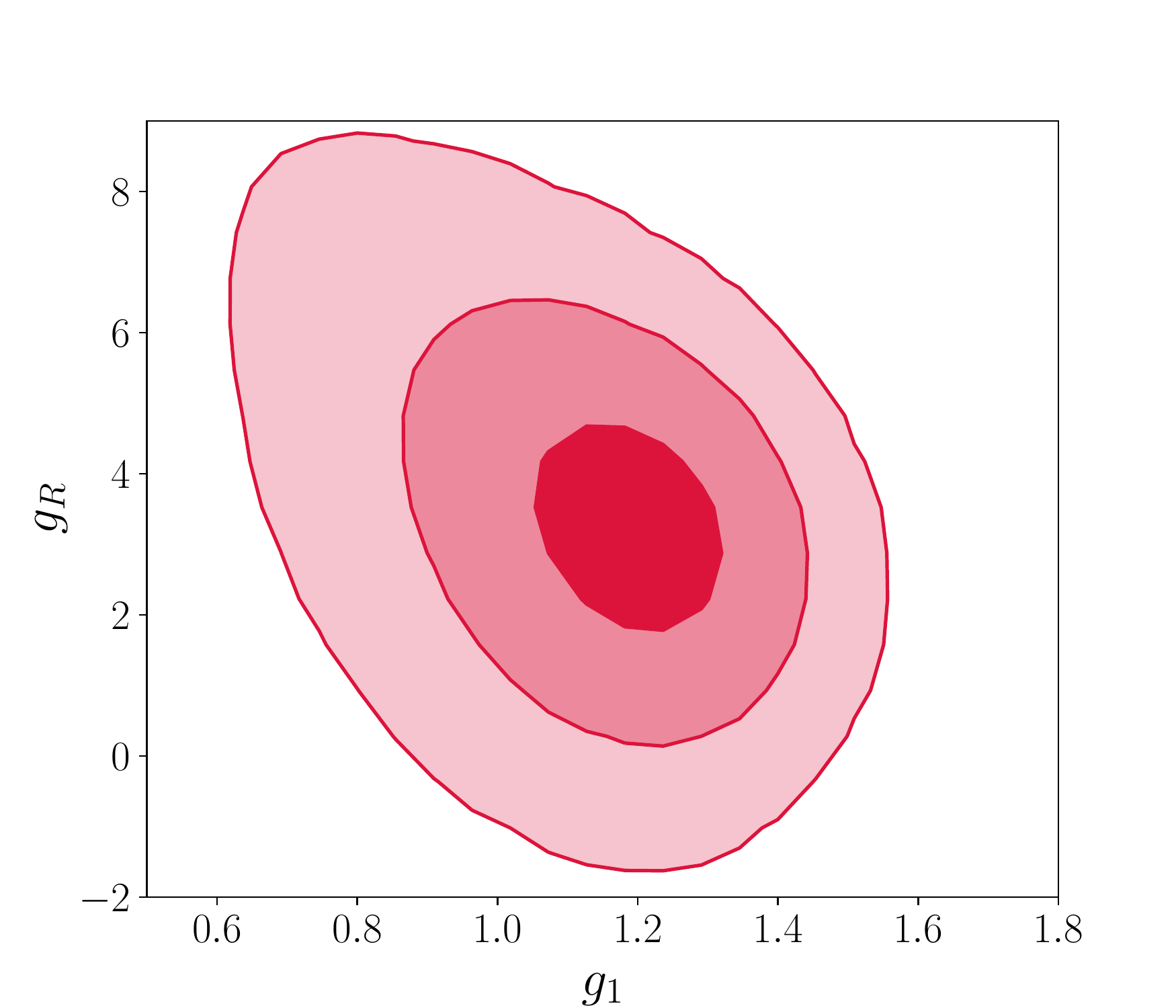}\hskip 5pt
\includegraphics[scale=0.45]{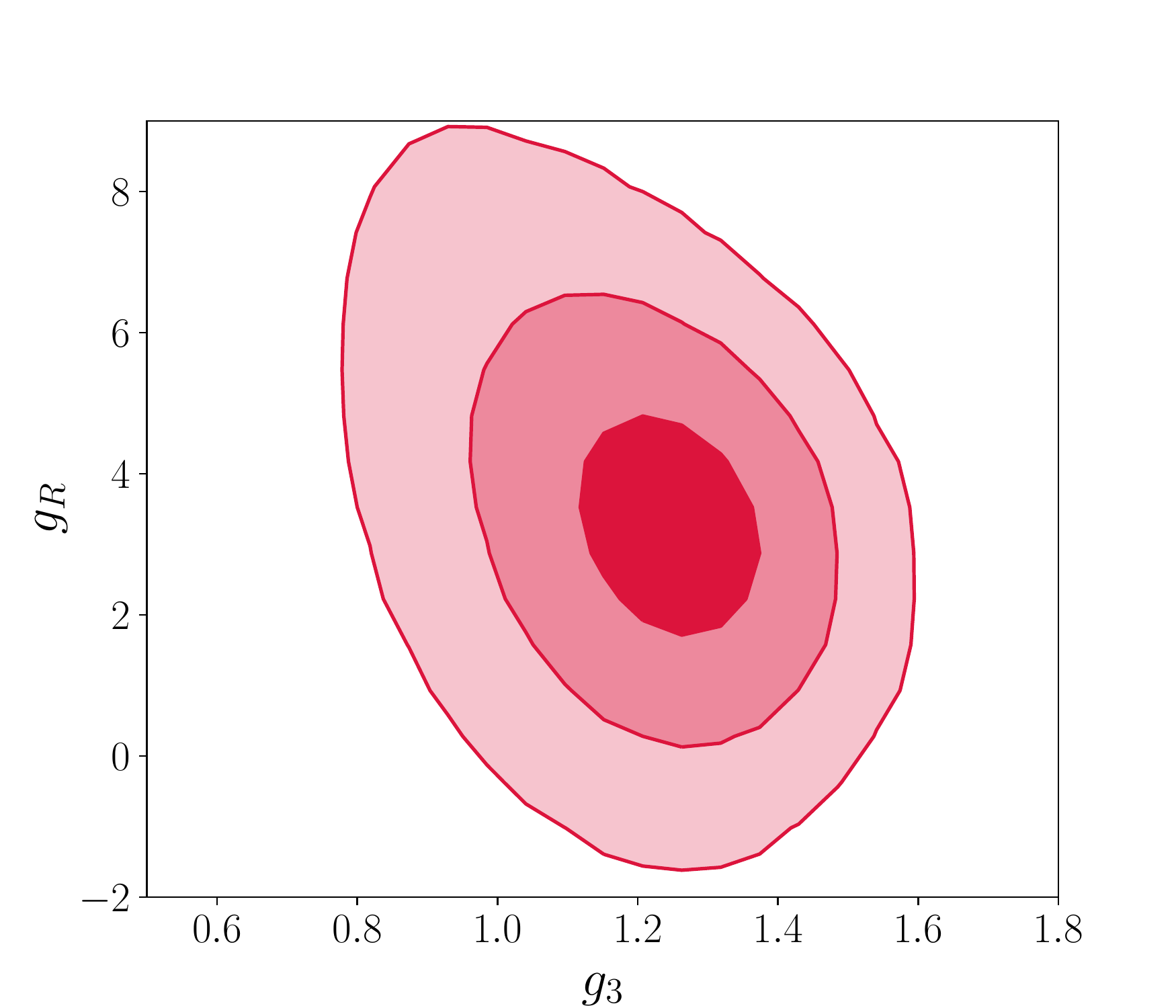}
\label{fig:contRH}
\caption{Fit results for the parameters $g_1$, $g_3$ and $g_R$. The shaded areas, from darker to lighter red, describe the $1$, $2$ and $3 \sigma$ regions, respectively.} 
\end{figure}

We now let $g_R\neq 0$. In this case there are two minima, corresponding to the two possibilities of satisfying $g_1 g_R>0$. For definiteness, we choose the case $g_1>0$ and $g_R>0$. We impose flat priors for $g_1$, $g_3$ and $g_R$ and obtain the following mode and $68\%$ interval for our fit parameters:
\begin{equation}
\begin{aligned}
g_1:\quad\text{mode}  = 1.23 \quad 68\%\,\, \text{interval}:[1.02,1.28] \,,\\
g_3:\quad\text{mode}  = 1.29 \quad 68\%\,\, \text{interval}:[1.10,1.33] \,,\\
g_R:\quad\text{mode}  = 3.13 \quad 68\%\,\, \text{interval}:[1.96,4.77]\,,
\end{aligned}    
\end{equation}
with correlation coefficients $\rho_{g_1g_3} = 0.92$, $\rho_{g_1g_R} = -0.40$ and $\rho_{g_3g_R} = -0.37$. In \fig{fig:contRH} we report the different two-dimensional projections. In this case we find $\Delta\chi^2 \sim 36$, which improves the fit of Benchmark I. We notice that our conclusion concerning the correlations between $g_1$ and $g_3$ remains unchanged, since the features leading to this result do not depend on right-handed currents. As expected, the addition of right-handed interactions only improves the fit, mostly by increasing the value of $R_{D^{(*)}}$ at the price of having moderate tensions with $\mu\to e\gamma$ and $\tau\to\mu\gamma$.

Concerning the observables of interest, we use the posteria distribution to analyse them. In the following we comment on the most interesting ones.

\paragraph{Charged currents:} Regarding $R_{D^{(*)}}$, we find
\begin{equation}
    R_D = 0.357 \pm 0.022 \,, \qquad R_{D^*} = 0.277 \pm 0.010\,,
\end{equation}
which correspond to an enhancement of $\sim 19\%$ and $\sim 9\%$ with respect to the SM predictions, respectively. 
Our predictions fall into the $1\sigma$ region, considerably easing the tension in these ratios.\footnote{We notice that the tensor coupling $\mathcal{F}^T_{D^{*}}(\tau)$ (see  \eq{eq:observable: RDst}) happens to be numerically sizeable, especially for $R_{D^{*}}$. This is however a quadratic NP contribution, and accordingly one expects it to be strongly suppressed. We have checked that this is the case.} However, for the longitudinal polarisation fraction $F_L^{D^*}$, we barely see an enhancement compared to the SM prediction. This is in line with what was already observed in previous analysis, as in refs.~\cite{Mandal:2020htr,Murgui:2019czp,Alok:2016qyh,Blanke:2018yud}.

Concerning additional modes mediated by $b\to c\tau\bar\nu$ transitions, an interesting observable is $R_{J/\psi} = \mathcal{B}(B_c\to J/\psi \tau\bar\nu)/\mathcal{B}(B_c\to J/\psi \mu\bar\nu)$. There is currently a discrepancy between the experimental measurement, $R_{J/\psi}^\text{exp} = 0.71\pm 0.17\text{(stat)}\pm 0.18\text{(syst)}$~\cite{Aaij:2017tyk} and the SM prediction from the lattice,  $R_{J/\psi}^\text{SM} = 0.2601\pm0.0036$~\cite{Harrison:2020nrv}.

For our estimate we use the lattice determination of the form factors in ref.~\cite{Harrison:2020gvo} and the predictions in ref.~\cite{Harrison:2020nrv}. Unfortunately, the lattice determination includes only SM form factors, while in our scenario sizeable tensor operators are generated. In order to estimate their impact, we assume that the ratio between scalar and tensor operators in $R_{J/\psi}$ is the same as in $R_{D^{*}}$. This yields:
\begin{equation}
\begin{aligned}
    R_{J/\psi}/R_{J/\psi}^\text{SM} &\approx |1+\mathcal{C}_L^{2333}|^2 - 0.095\, \Re[(1+\mathcal{C}_L^{2333})\mathcal{C}_S^{* 2333}]+0.034\, |\mathcal{C}_S^{2333}|^2\\
    &- 4.255\,  \Re[(1+\mathcal{C}_L^{2333})\mathcal{C}_T^{*2333}]+16.183\, |\mathcal{C}_T^{2333}|^2\,.
    \end{aligned}
    \label{eq:RJpsi}
\end{equation}
The full basis of form factors for $B_c\to J/\psi \ell\bar\nu$ decays has been determined in ref.~\cite{Leljak:2019eyw} using Light-Cone Sum Rules. However, this determination has large uncertainties, especially for the tensor form factors. We have checked that \eq{eq:RJpsi} shows numerical agreement with the results in ref.~\cite{Leljak:2019eyw}. Our prediction for $R_{J/\psi}$ reads:
\begin{equation}
    R_{J/\psi} = 0.279\pm 0.007\,,
\end{equation}
which is very close to the lattice result. Thus, our framework barely reduces the present tension in $R_{J/\psi}$.

Another interesting mode is $B_c\to\tau\bar\nu$. In our setting, the dominant NP contribution comes from the interference between SM-like and scalar operators, which happens to be destructive, and we find $\mathcal{B}(B_c\to\tau\bar\nu)\sim 0.7\%$, while the SM prediction is $\sim 2\%$. We note that this sizeable reduction of the SM estimate does not collide with constraints from the $B_c$ lifetime \cite{Alonso:2016oyd}.   

\paragraph{Neutral meson mixing:} The bounds from neutral $K$ and neutral $D$ mixing are easily satisfied in our framework. Concerning $\Delta M_s$ and $\Delta M_d$, the situation is rather different. In the SM there is already a tension between the theory determinations and experimental measurements of about $0.8\sigma$ and $0.4\sigma$, respectively~\cite{DiLuzio:2019jyq}. In our framework, these tensions increase to $1.4\sigma$ and $0.8\sigma$, respectively. 

\paragraph{FCNCs:} Right-handed interactions introduce potentially sizeable effects in $D$ decays through chirally enhanced scalar operators. We have checked that our predictions are well below the current bounds for leptonic $D$ decays.

Concerning FCNCs in the down-type quark sector, our predictions for $B^+\to K^+\tau^+\tau^-$ and  $\bar{B}_s\to\tau^+\tau^-$ decays are one order of magnitude enhanced with respect to their SM prediction. This feature is common to many NP scenarios, since the couplings to third generation fermions are enhanced by the need to match the discrepancies observed in $R_{D^{(*)}}$. The current experimental limits are still orders of magnitude above our predictions (see \Table{tab:prediction}), but prospects for future measurements at LHCb and Belle II render these decay modes suitable tests of physics beyond the SM.

Regarding LFV $B$ decays, we obtain a sizeable enhancement in $b\to s\tau^\pm\mu^\mp$ transitions, while all the other decay modes are negligible. In particular, in \Table{tab:prediction} we provide predictions for $\bar{B}_s\to \tau^\pm \mu^\mp$ and $B^+\to K^+\tau^+\mu^-$,\footnote{Our solutions predict $B^+\to K^+\tau^-\mu^+$ to be parametrically smaller than $B^+\to K^+\tau^+\mu^-$, and accordingly the constraints for the former mode are less significant.} where for the latter we use the form factors from ref.~\cite{Bouchard:2013eph}. These modes will be tested with increasing precision at LHCb and Belle II. 

We have also examined the invisible decay modes $K^+\to\pi^+\nu\bar\nu$ and $B\to K^{(*)}\nu\bar\nu$. Our predictions are very close to the most recent experimental limits. Interestingly, in both cases the LFV contributions are as sizeable as the LFC ones. Accordingly, both contributions have been included in the fit.

\paragraph{LFV $\tau$ decays:} These are the most striking probes of our framework. The branching fraction of $\tau\to\mu\phi$ is predicted to be roughly an order of magnitude larger than the current experimental limit (see \Table{tab:prediction}). The most sensitive mode is $\tau\to\mu\gamma$. Its branching fraction is dominated by top-quark loops, in which the left- and right-handed contributions interfere. Numerically, we obtain
\begin{equation}
    \mathcal{B}(\tau\to\mu\gamma) \sim 8.67\times  10^{-7} g_1^2 g_R^2 (c_L^{32})^2 (c_R^{33})^2\,.
\end{equation}
Given the asymmetric distribution of the fitted parameters $g_1$ and $g_R$, the distribution of $ \mathcal{B}(\tau\to\mu\gamma)$ is also highly asymmetric. We find the following 68\% interval:
\begin{equation}
  \frac{\mathcal{B}(\tau\to\mu\gamma)}{(c_L^{32})^2 (c_R^{33})^2} \in [0.420, 2.38]\times 10^{-5}\,.
\end{equation}
As already discussed in Sect.~\ref{sec:4}, the tensions in $R_{D^{(*)}}$ entail that the current experimental limit for $\tau\to\mu\gamma$ (see \eq{eq:datalgamma3}) is overshooted, from power-counting considerations only. This can be resolved, e.g., by nominally choosing 
\begin{align}\label{reduced}
c_L^{32} = -\frac{1}{4}\quad {\textrm{and}}\quad  c_R^{33} = \frac{1}{4}\,,
\end{align}
such that
\begin{equation}
     \mathcal{B}(\tau\to\mu\gamma) \in [1.78, 9.52]\times 10^{-8}\,,
\end{equation}
which now largely overlaps with the current experimental limits. We stress that these reduction of the Wilson coefficients only affects the prediction for $\tau\to\mu\gamma$. In particular, it has no effect on any of our predictions, i.e.\ those in \Table{tab:prediction} and the one for $R_{J/\psi}$. 

The situation is qualitatively similar for $\tau\to 3\mu$, where the main contribution comes from the decay chain $\tau\to\mu\gamma(\to\mu^+\mu^-)$. However, the current experimental bounds are slightly weaker than the ones in $\tau\to\mu\gamma$ and they are easily satisfied.

\begin{table}
\begin{center}
\renewcommand{\arraystretch}{1.2} 
\begin{tabular}{l  c c c}
\toprule
 Mode & $\mathcal{B}_{\rm SM} $  & $\mathcal{B}_{\rm Exp}$  & This work  \\
\midrule
$\tau \to \mu \phi $ & 0 & $< 8.4 \times 10^{-8}$~\cite{Tanabashi:2018oca} & $[0.583,1.25]\times 10^{-10}$ \\
$B_s \to \tau^\pm \mu^\mp $ & 0 & $< 4.2 \times 10^{-5}$~\cite{Aaij:2019okb} & $[1.21,2.60]\times 10^{-6}$ \\
$B^+ \to K^+ \tau^+ \tau^- $ & $\left( 1.60 \pm 0.12 \right)\times 10^{-7} $~\cite{Du:2015tda}  & $< 2.2 \times 10^{-3}$~\cite{TheBaBar:2016xwe} & $[7.87,13.3]\times \mathcal{B}_{\rm SM}$  \\
$\bar B_s \to\tau^+ \tau^- $ & $\left(7.30 \pm 0.49 \right)\times 10^{-7} $~\cite{Bobeth:2013uxa}  & $< 6.8 \times 10^{-3}$~\cite{Aaij:2017xqt} & $[7.75,13.1]\times \mathcal{B}_{\rm SM}$  \\
$B^+ \to K^+ \tau^+ \mu^- $ & 0 &  $< 3.9 \times 10^{-5}$~\cite{Aaij:2020mqb}  & $\left( 1.8 \pm 0.7\right)\times 10^{-6}$  \\
\toprule
\end{tabular}
\caption{Predictions of the branching fractions for several decay modes. The quoted experimental upper limits are at 90\% C.L.. }
\label{tab:prediction}
\end{center}
\end{table}

\paragraph{Radiative muon modes:} The $S_1$ leptoquark contributes to the muon $g-2$. In our framework, the major contribution arises from the combination of left- and right-handed $S_1$ leptoquark spurions with a top quark in the loop. Using  \eq{eq:g-2}, we find
\bea
\Delta a_\mu \simeq -3.67 \times 10^{-7} \Re[g_1 g_R^* S_{QL}^{32} S_{UE}^{*32}]\,,
\eea
which requires $g_1 g_R^* S_{QL}^{32} S_{UE}^{*32}\sim \lambda^3$ to accommodate the current experimental measurement. Scenario A yields an extremely suppressed contribution. However, in Scenario B we find 
\bea
\Delta a_\mu \sim 1.1 \times10^{-10} c_R^{*32} c_L^{32}\,,
\eea
where the size of $c_L^{32}$ is fixed by $\tau\to\mu\gamma$. It is not easy to describe the two observables simultaneously. If $c_R^{32}$ is set to $1$ and we include $\Delta a_\mu$ in our fit, we find a solution where a slightly higher value of $g_R$ is preferred. However, the description of $\tau\to\mu\gamma$ worsens and $\Delta a_\mu$ still deviates from its SM prediction by more than $3\sigma$. Only imposing  $c_R^{32}\sim\mathcal{O}(10)$ one can reproduce the experimental measurement. However, this large coefficient is barely compatible with the power counting. We therefore do not include $\Delta a_\mu$ in our fit.

Regarding the $\mu\to e\gamma$ mode, it is naturally suppressed in Scenario B. Instead, in Scenario A we have:
\begin{equation}
    \mathcal{B}(\mu\to e\gamma)\in [0.48,2.5]\times 10^{-13}\,,
\end{equation}
where the upper and lower extrema encode the 68\% interval. This prediction is very close to the current experimental upper limits. Notice that in this case the choice of $b_L^1 =-5$ is necessary to suppress the enhancement due to $g_R$. Future measurements of $\mathcal{B}(\mu\to e\gamma)$ at the MEG II experiment will improve the current limits.
Notice that in our scenario limits on $\mu\to e \gamma$ provide a bound on $b_L^1$. Hence, in our setup, prospective more stringent upper limits on this mode can be easily accommodated.

\subsection{Comparison with related literature}

The results of our analysis share a number of features worth comparing with recent related works. In particular, it is instructive to compare with ref.~\cite{Bordone:2019uzc}, where the same power counting scheme was used for the vector leptoquark $U_1$. Interestingly, both vector and scalar leptoquarks lead to essentially the same constraints for the FN charges of the left-handed fermion sector. Phenomenologically, in both cases the left-handed couplings are not enough to explain the tensions in $R_{D^{(*)}}$ and require the contribution of the right-handed sector. However, whereas the $U_1$ couples to the right-handed down-quark sector, the $S_1$ affects processes sensitive to the right-handed up-quark sector. As a result, the FN charges in the $U_1$ case are mostly constrained from both lepton-flavour conserving and lepton-flavour violating FCNCs, such as $\bar B_s\to\tau^+\tau^-$ and $\bar B_{s,d}\to \tau^\pm\mu^\mp$. Instead, in the $S_1$+$S_3$ scenario, the contributions from radiative leptonic LFV decays, mostly $\tau\to \mu\gamma$ and $\mu\to e \gamma$, are the driving force to fix the FN charges for the right-handed sector. The result is that the number of phenomenologically allowed solutions in the scalar leptoquark scenario is substantially small and leads to a qualitatively similar set of predictions. This is in contrast with the $U_1$ case, where the different solutions lead to rather distinct phenomenological scenarios.   

It is also relevant to compare our analysis with other studies of the $S_1$+$S_3$ scenario, in particular with the recent analysis of ref.~\cite{Gherardi:2020qhc}, which contains a set of phenomenological observables comparable to the ones used in this paper. The strategy in ref.~\cite{Gherardi:2020qhc} is to fit the leptoquark flavour couplings to data, with the freedom to set to zero certain couplings in order to fulfill phenomenological constraints. This is in contrast with our approach, where the power-counting structure sets a more rigid framework. In particular, we do not suppress spurion entries by sending couplings to zero. It is interesting to remark that with only left-handed interactions, the results of ref.~\cite{Gherardi:2020qhc} are compatible with our FN power counting for the flavour couplings. With the addition of right-handed interactions, this is no longer the case.   

The main phenomenological difference between ref.~\cite{Gherardi:2020qhc} and the present work is the prediction for the muon $(g-2)$. As already discussed, with our power counting and a reference leptoquark mass of $M=2$ TeV, a sizable contribution to the muon $(g-2)$ increases the tension between $\tau\to\mu\gamma$ and $R_{D^{(*)}}$, which can only be resolved with a rather unnatural enhancement of certain Wilson coefficients. These tensions are partially alleviated in ref.~\cite{Gherardi:2020qhc} by having lighter leptoquarks and a larger set of free parameters.

\section{Conclusions}
\label{sec:6}

One manifestation of the flavour problem in the Standard Model is the number of free parameters in the Yukawa couplings of fermions to the Higgs boson. When effective field theories are employed to incorporate new-physics effects, the problem gets amplified and the number of flavour-specific parameters soon becomes
intractable. A framework to reduce this complexity in an effective field theory-inspired way is to work with selected flavour 
spurions and provide them with a power counting based on Froggatt-Nielsen charges, as suggested in ref.~\cite{Bordone:2019uzc}. Given the tensions currently observed in various $B$-meson decays, natural choices
of flavour spurions are those linked to the specific new-physics scenarios that can address these tensions. The Froggatt-Nielsen power counting can then lead to deeper insights into the new physics flavour patterns observed in the low-energy data.

In this paper we single out the flavour spurions associated with the scalar leptoquarks $S_1$ and $S_3$.
Compliance with low-energy experimental constraints reduces the allowed scenarios for the Froggatt-Nielsen charges, in the minimal setup we adopt, to essentially two. Both of them predict values for the branching ratios of $\tau\to\mu\gamma$, $\mu\to e\gamma$, $\bar B_s\to\tau^\pm\mu^\mp$ and $B^+\to K^+\tau^+\mu^-$ close to the current experimental limits. Interestingly, we also predict a decrease in $B_c\to\tau{\bar{\nu}}$ to one third of the SM estimate. 

We also conclude, like previous studies, that both $S_1$ and $S_3$ are needed in order to match the experimental values of $R_{D^{(*)}}$, with a relevant role played by the scalar and tensor operators stemming from the $S_1$ interactions to right-handed fermions. The current excess in $R_{D^{(*)}}$, together with the experimental measurements of e.g. $K\to\pi\nu{\bar{\nu}}$, $B\to K^*\nu{\bar{\nu}}$, 
neutral meson mixing (mostly $B_s$ and $B_d$), $Z\to \nu {\bar{\nu}}$, $Z\to\tau^+\tau^-$, $\mu\to e \gamma$ and 
$\tau\to\mu\gamma$ constrain the Froggatt-Nielsen charges and generate phenomenologically interesting flavour patterns. The scalar-leptoquark scenario is a renormalizable extension of the SM and, as such, allows one to perform a complete analysis of both tree-level and loop-induced constraints. This increase in the number of constraints reduces the size of the allowed parameter space of the Froggatt-Nielsen charges with respect to the $U_1$ case~\cite{Bordone:2019uzc}.

Finally, we find that our flavour structure does not allow for a simultaneous account of $R_{D^{(*)}}$, the anomalous magnetic moment of the muon $(g-2)$ and the branching ratio of $\tau\to\mu\gamma$, unless some of the  Wilson coefficients in front of the spurion couplings can take large values, beyond the power-counting expectation. Only the discussion of a UV realisation of our framework can assess whether such deviations from the power counting are to be expected or not.

\section*{Acknowledgements}
We thank Adam Falkowski, Alexander Lenz and David Marzocca for useful communications. M.B. is grateful to Ayan Paul for helpful discussions. The work of M.B., O.C., T.F. is supported by Deutsche Forschungsgemeinschaft (DFG, German Research Foundation) under grant 396021762 - TRR 257 ``Particle Physics Phenomenology after the Higgs Discovery''. The work of M.B. is partly supported by the Italian Ministry of Research (MIUR) under grant PRIN 20172LNEEZ.
The work of R.M. is supported by the Alexander von Humboldt Foundation through a postdoctoral research fellowship.
\appendix

\section{Matching to SMEFT and LEFT}
At energies below the leptoquark masses and above the electroweak scale, a suitable EFT description for our framework is the SMEFT, where the leptoquarks are integrated out. We choose the so-called Warsaw basis in ref.~\cite{Grzadkowski:2010es} and we obtain:
\begin{equation}
\begin{aligned}
\mathcal{L}_\text{eff}=\mathcal{L}_\text{SM} - \frac{1}{M^2}\bigg\{[&\mathcal{C}_{lq}^{(3)}]^{ij\alpha\beta}(\bar{Q}^i \gamma^\mu \sigma^a Q^j)(\bar{L}^\alpha \gamma_\mu \sigma^a L^\beta)+[\mathcal{C}_{lq}^{(1)}]^{ij\alpha\beta}(\bar{Q}^i \gamma^\mu Q^j)(\bar{L}^\alpha \gamma_\mu  L^\beta) \\
+\,&[\mathcal{C}_{eu}]^{ij\alpha\beta}(\bar{u}^i_R\gamma^\mu u_R^j)(\bar{e}^\alpha_R\gamma_\mu e^\beta_R)+[\mathcal{C}_{lequ}^{(1)}]^{ij\alpha\beta}(\bar{Q}^i u_R^j)\epsilon(\bar{L}^\alpha e^\beta_R) \\
+\,&[\mathcal{C}_{lequ}^{(3)}]^{ij\alpha\beta}(\bar{Q}^i \sigma^{\mu\nu}u_R^j)\epsilon(\bar{L}^\alpha\sigma_{\mu\nu} e^\beta_R)  \bigg\} \,.
\end{aligned}
\label{eq:lagrangian_SMEFT}
\end{equation}
At energies below the electroweak scale, it is more convenient to use the Low Energy Effective Lagrangian (LEFT), which can be found in ref.~\cite{Jenkins:2017jig}. Our scenario yields
\begin{equation*}
    \mathcal{L}_\text{eff}=\mathcal{L}_\text{SM} -\frac{1}{M^2} \sum_i \left[L_i \mathcal{O}_i + \text{h.c.}\right]\,,
\end{equation*}
where
\begin{equation}
\begin{aligned}
\left[\mathcal{O}_{\nu d}^{V,LL}\right]^{ij\alpha\beta} =&\,(\bar{d}_L^i\gamma^\mu d_L^j)(\bar{\nu}_L^\alpha\gamma_\mu \nu_L^\beta) \,, & \left[\mathcal{O}_{eu}^{V,LL}\right]^{ij\alpha\beta} =&\,(\bar{u}_L^i\gamma^\mu u_L^j)(\bar{e}_L^\alpha\gamma_\mu e_L^\beta) \,,  \\
\left[\mathcal{O}_{e d}^{V,LL}\right]^{ij\alpha\beta}=&\,(\bar{d}_L^i\gamma^\mu d_L^j)(\bar{e}_L^\alpha\gamma_\mu e_L^\beta) \,, & \left[\mathcal{O}_{\nu e du}^{V,LL}\right]^{ij\alpha\beta}=&\,(\bar{d}_L^i\gamma^\mu u_L^j)(\bar{\nu}_L^\alpha\gamma_\mu e_L^\beta)\\
\left[\mathcal{O}_{eu}^{V,RR}\right]^{ij\alpha\beta} =&\,(\bar{u}_R^i\gamma^\mu u_R^j)(\bar{e}_R^\alpha\gamma_\mu e_R^\beta) \,,  & 
[\mathcal{O}_{\nu edu}^{S,RR}]^{ij\alpha\beta} =&\,  (\bar{d}^i_L u^j_R)(\bar\nu^\alpha_L e^\beta_R) \,, \\ [\mathcal{O}_{\nu e d u}^{T,RR}]^{ij\alpha\beta} =&\,  (\bar{d}_L^i \sigma^{\mu\nu}u_R^j)(\bar\nu_L^\alpha\sigma_{\mu\nu} e_R^\beta) \,,  &
[\mathcal{O}_{eu}^{S,RR}]^{ij\alpha\beta} =&\,(\bar{u}_L^i u_R^j) (\bar{e}_L^\alpha\beta e_R) \,, \\ [\mathcal{O}_{e u}^{T,RR}]^{ij\alpha\beta} =&\, (\bar{u}^i_L \sigma^{\mu\nu} u_R^j)(\bar{e}_L^\alpha \sigma_{\mu\nu} e_R^\beta)\,.  
\end{aligned}
\end{equation}
The Wilson coefficients of the LEFT Lagrangian can be written in terms of the SMEFT Wilson coefficients as 
\begin{equation}
    \begin{aligned}
    \big[L_{\nu d}^{V,LL}\big]^{ij\alpha\beta} =  [L_{eu }^{V,LL}]^{ij\alpha\beta} =& \, [\mathcal{C}_{lq}^{(1)}]^{ij\alpha\beta}-[\mathcal{C}_{lq}^{(3)}]^{ij\alpha\beta}\,, \\
    [L_{ed}^{V,LL}]^{ij\alpha\beta} =& \, [\mathcal{C}_{lq}^{(1)}]^{ij\alpha\beta}+[\mathcal{C}_{lq}^{(3)}]^{ij\alpha\beta} \,,\\
    [L_{\nu edu}^{V,LL}]^{ij\alpha\beta} =&\, 2 [\mathcal{C}_{lq}^{(3)}]^{ij\alpha\beta}\,, \\
    [L_{eu}^{V,RR}]^{ij\alpha\beta} =& \, [\mathcal{C}_{eu}]^{ij\alpha\beta} \,,\\
    [L_{\nu edu}^{S,RR}]^{ij\alpha\beta} = -[L_{ eu}^{S,RR}]^{ij\alpha\beta} = & \, [\mathcal{C}_{lequ}^{(1)}]^{ij\alpha\beta} \,,\\
    [L_{\nu edu}^{T,RR}]^{ij\alpha\beta} = -[L_{ eu}^{T,RR}]^{ij\alpha\beta} = & \, [\mathcal{C}_{lequ}^{(3)}]^{ij\alpha\beta} \,.\\
    \end{aligned}
\end{equation}
For convenience, throughout this paper we have expressed the NP contribution to the Wilson coefficients in terms of the SMEFT coefficients. The RGE running from the electroweak to the hadronic scale has been considered for the scalar and tensor operators entering $R_{D^{(*)}}$, which is the only relevant one for the observables under study (see Appendix~\ref{app:B}).
\section{Observables}
\label{app:B}
\subsection{$d_j\to u_i\ell_\alpha\bar\nu_\beta$}
\label{app:B1}

The charged-current transitions, $d_j\to u_i \ell_\alpha\bar\nu_\beta$, are described by the following Lagrangian:
\be
\label{eq:btoc}
\begin{aligned}
\mathcal{L}(d_j\to u_i \ell_\alpha \bar\nu_\beta)=-\frac{4G_F}{\sqrt{2}} V_{ij} &\left[(\delta_{\alpha\beta}+\mathcal{C}_L^{ij\alpha\beta})(\bar{u}^i\gamma^\mu P_L d^j)(\bar{e}^\alpha\gamma_\mu P_L \nu^\beta) + \mathcal{C}_S^{ij\alpha\beta} (\bar{u}^i P_Ld^j)(\bar{e}^\alpha P_L\nu^\beta) \right.\\
&\left.+\,\mathcal{C}_T^{ij\alpha\beta} (\bar{u}^i \sigma^{\mu\nu} P_L d)(\bar{e}^i\sigma_{\mu\nu}P_L\nu^\beta)\right ] \,.
\end{aligned}
\ee
The NP Wilson coefficients read
\begin{align}
\mathcal{C}_{L}^{ij\alpha\beta}=&+ \frac{v^2}{M^2}\, \sum_{m=1}^3 \frac{V_{im}}{V_{ij}} [\mathcal{C}^{(3)}_{lq}]^{mj\alpha\beta}  \,,  \label{eq:btoc_coeffs1}\\
\mathcal{C}_{S}^{ij\alpha\beta} =&-\frac{v^2}{2 M^2 V_{ij}}\,[\mathcal{C}^{(1)*}_{lequ}]^{ij\alpha\beta}\,. \label{eq:btoc_coeffs2}  \\
\mathcal{C}_{T}^{ij\alpha\beta}=&-\frac{v^2}{2 M^2 V_{ij}} [\mathcal{C}^{(3)*}_{lequ}]^{ij\alpha\beta}\,. \label{eq:btoc_coeffs3} 
\end{align}

We note that the matching applies at the high scale $M$ and we need to evolve these predictions, using the renormalization group, to the much lower scales where hadronic decays take place. Neglecting electroweak corrections, we write for scalar and tensor operators 

\bea
\label{eq:running}
C _{S,T} (\mu)\, =\, \left(\frac{\alpha_s^{(n_f)}(\mu)}{\alpha_s^{(n_f)}(m_q^{f+1})}\right)^{-\frac{\gamma_{S,T}}{2\beta_0^{(n_f)}}}
\cdots\quad
\left(\frac{\alpha_s^{(5)}(m_b)}{\alpha^{(5)}_s(m_t)}\right)^{-\frac{\gamma_{S,T}}{2\beta_0^{(5)}}} \;
\left(\frac{\alpha_s^{(6)}(m_t)}{\alpha^{(6)}_s(M)}\right)^{-\frac{\gamma_{S,T}}{2\beta_0^{(6)}}} C_{S,T}(M) \,,
\eea
where the anomalous dimensions are  $\gamma_{S}=-8$, $\gamma_{T}=-8/3$, respectively~\cite{Aebischer:2017gaw} and the leading term in the QCD beta function is given by $\beta_0^{(n_f)}= 11- 2\, n_f /3 $.

At the $m_b$-scale, we find
\bea
\mathcal{C}_S(m_b) = 1.67\, \mathcal{C}_S (M)~~{\rm and}~~ \mathcal{C}_T(m_b) = 0.84\, \mathcal{C}_T (M)\,.
\eea

The most interesting channels for testing the $b\to c$ transitions are $B$ meson decays. The observables driving the NP effects are the universality ratios $R_{D^{(*)}}$.  In order to evaluate the NP effects in presence of non SM interactions, we refer to the full kinematical distributions for $B\to D^{(*)}$ semileptonic decay in ref.~\cite{Murgui:2019czp}. After integrating the kinematical distributions, we derive the following expressions for the universality ratios $R_{D^{(*)}}$:
\begin{equation}
\label{eq:observable: RDst}
\begin{aligned}
R_{D^{(*)}}\approx& R_{D^{(*)}}^{\text{SM}} \left\{ |1+ \mathcal{C}_L^{2333}|^2+\mathcal{F}^S_{D^{(*)}}(\tau)\, |\mathcal{C}_{S}^{2333}|^2+\mathcal{F}^T_{D^{(*)}}(\tau)|\mathcal{C}_{T}^{2333}|^2 \right. \\
&\qquad \left. + \mathcal{F}^{VS}_{D^{(*)}}(\tau) \, \Re[(1+ \mathcal{C}_L^{2333})\,\mathcal{C}_{S}^{*2333}] +  \mathcal{F}^{VT}_{D^{(*)}} (\tau)\,\Re[(1+ \mathcal{C}_L^{2333})\,\mathcal{C}_{T}^{*2333} ]\right\} \,,
\end{aligned}
\end{equation}
where the functions $\mathcal{F}^{S(T)}_{D^{(*)}}(\tau)$ are a placeholder for the integrals over kinematics and form factors associated with the scalar (tensor) contributions for a $D$ or a $D^*$ meson. Note that the quantities $\mathcal{F}^{S(T)}_{D^{(*)}}(\mu)\sim 0$, since they are suppressed by the muon mass and hence are neglected. We adopt the values from ref.
~\cite{Mandal:2020htr}, namely  $\mathcal{F}^{S}_{D^{(*)}}(\tau) =1.037 (0.037) $, $\mathcal{F}^{T}_{D^{(*)}}(\tau) =0.939 (17.378)$, $\mathcal{F}^{VS}_{D^{(*)}}(\tau) = 1.504 (-0.114)$ and  $\mathcal{F}^{VT}_{D^{(*)}}(\tau) = 1.171 (-5.130)$. We stress that these values are almost independent of the form factor parametrisation used, and they largely agree with ref.\cite{Fajfer:2012vx,Bordone:2019guc}. Furthermore, we neglect possible LFV contributions. The values used in the fit are reported in \Table{tab:LFU_tau}. Concerning $R_{D^{(*)}}$ we perform the arithmetic mean of the values in refs.\cite{Bigi:2016mdz,Bernlochner:2017jka,Jaiswal:2017rve,Bordone:2019guc}. \\
Another interesting measurement of the longitudinal polarization fraction for $D^*$ in the $B \to D^* \tau \bar{\nu}$ mode has been recently performed by Belle~\cite{Abdesselam:2019wbt}, where $ F_L^{D^*}|_{\rm exp}=0.60 \pm 0.08 \pm 0.035$ lies $1.7\sigma$ above the SM expectation, which is around $0.45$~\cite{Bhattacharya:2018kig,Bordone:2019guc}. The corresponding expression including the NP operators can be written as~\cite{Mandal:2020htr}
\begin{equation}
\label{eq:observable:FLDst}
\begin{aligned}
F_L^{D^{*}}\approx&\, R_{D^{(*)}}^{-1} \left\{ 0.120 |1+ \mathcal{C}_L^{2333}|^2+ 0.010\, |\mathcal{C}_{S}^{2333}|^2+ 0.869 |\mathcal{C}_{T}^{2333}|^2 \right. \\
&\qquad \left. -0.030 \, \Re[(1+ \mathcal{C}_L^{2333})\,\mathcal{C}_{S}^{*2333}] -0.525\,\Re[(1+ \mathcal{C}_L^{2333})\,\mathcal{C}_{T}^{*2333} ]\right\} \,.
\end{aligned}
\end{equation}

To understand if and how well universality holds for decays into light leptons, one can compare $|V_{cb}|$ as extracted from electron and muon modes. If we define  $|\tilde{V}_{cb}^\ell|$ as the effective $|V_{cb}|$ in the presence of NP contributions associated with a lepton $\ell$, the universality between $\mu$ and $e$ is measured by
\be
\frac{|\tilde{V}^e_{cb}|}{|\tilde{V}^\mu_{cb}|}=\left[ \frac{\vert1+\mathcal{C}_L^{2311}\vert^2+|\mathcal{C}_L^{2321}|^2+\vert\mathcal{C}_L^{2331}|^2}{\vert1+\mathcal{C}_L^{2322}\vert^2+|\mathcal{C}_L^{2312}|^2+|\mathcal{C}_L^{2332}|^2}\right]^{\frac{1}{2}} \,.
\ee
Contributions from scalar and tensor operators are suppressed by the light lepton masses and can be safely neglected. The numerical inputs used in the fit are reported in \Table{tab:LFU_tau}.\\

Finally, it is also interesting to consider the leptonic decay modes of charged $B_q$ mesons, where $q=u,c$. The corresponding branching ratio reads
\begin{align}
\mathcal{B}(B_q\to \ell\bar\nu) &=\mathcal{B}(B_q\to \ell\bar\nu)\vert_\text{SM}\left(\left|1+\mathcal{C}_L^{q3\ell\ell}+\frac{m_{B_q}^{2}}{m_\ell (m_b+m_q)}\mathcal{C}_{S}^{q3\ell\ell}\right|^2  \right.
\cr  & \qquad \phantom{\mathcal{B}(B_q\to \ell\bar\nu)\vert_\text{SM}} \left. +
\sum_{\ell\neq\ell^\prime}\left|\mathcal{C}_L^{q3\ell\ell^\prime}+\frac{m_{B_q}^{2}}{m_\ell (m_b+m_q)}\mathcal{C}_{S}^{q3\ell\ell^\prime}\right|^2 \right) \,.
\end{align}

\begin{table}
\begin{center}
\renewcommand{\arraystretch}{1.2} 
\begin{tabular}{c c c c}
\toprule
Observable & Measurement & Correlation & SM \\
\midrule
$R_D$	& $0.340 \pm0.027 \pm 0.013 $	&\multirow{2}{*}{-0.38} &  $0.299\pm 0.003$ \cite{Bigi:2016mdz,Bernlochner:2017jka,Jaiswal:2017rve,Bordone:2019guc,Amhis:2019ckw} \\
$R_{D^*}$ & $0.295 \pm 0.011 \pm 0.008 $	&		& $0.255\pm0.007$ \cite{Bigi:2017jbd,Bernlochner:2017jka,Jaiswal:2017rve,Bordone:2019guc,Amhis:2019ckw}\\
\midrule
$V_{cb}\vert_D$ & $1.004(42)$ \cite{Jung:2018lfu}& \multirow{2}{*}{-} & 1. \\
$V_{cb}\vert_{D^*}$ & $0.97(4)$ \cite{Jung:2018lfu} &  &1.\\
\toprule
\end{tabular}
\caption{Experimental measurements, SM predictions and correlations for $b\to c$ transitions.}
\label{tab:LFU_tau}
\end{center}
\end{table}

\subsection{$d_j\to d_i\ell_\alpha \ell_\beta$}
\label{app:dtodll}
The effective Lagrangian describing a generic $d_j\to d_i \ell_\alpha\ell_\beta$ FCNC transition reads
\be
\begin{aligned}
\mathcal{L}_\text{NP}(d_j\to d_i \ell_\alpha\ell_\beta)=\frac{4 G_F}{\sqrt{2}}\frac{\alpha_\text{EM}}{4\pi}V_{td_j}V^*_{td_i}&\left[(\mathcal{C}_9^\text{SM}\delta_{\alpha\beta}+\mathcal{C}_9^{ij\alpha\beta})\mathcal{O}_9^{ij\alpha\beta}\right. \\
+&\left.(\mathcal{C}_{10}^\text{SM}\delta_{\alpha\beta}+\mathcal{C}_{10}^{ij\alpha\beta})\mathcal{O}_{10}^{ij\alpha\beta} \right ]\,,
\end{aligned}
\label{eq:lagrangian_bsll}
\ee
where
\begin{align}
\mathcal{O}_9^{ij\alpha\beta} = & \,(\bar{d}_i \gamma^\mu P_L d_j)(\bar{\ell}_\alpha \gamma_\mu \ell_\beta)  \,, & \mathcal{O}_{10}^{ij\alpha\beta} = &\,(\bar{d}_i \gamma^\mu P_L d_j)(\bar{\ell}_\alpha \gamma_\mu\gamma_5 \ell_\beta) \,.
\label{eq:base_NC}
\end{align}
Using the results of the previous Appendix, the matching to SMEFT Wilson coefficients reads
\be
\begin{aligned}
\mathcal{C}_{9}^{ij\alpha\beta} = - \,\mathcal{C}_{10}^{ij\alpha\beta} =& -\frac{v^2}{M^2}\frac{\pi}{\alpha_\text{EM} V_{td_j}V^*_{td_i}} \left([\mathcal{C}_{lq}^{(3)}]^{ij\alpha\beta}+[\mathcal{C}_{lq}^{(1)}]^{ij\alpha\beta}\right)\,.
\end{aligned}
\label{eq:observables_bsll}
\ee
From the channels considered in the main text, the most stringent bounds come from measurements of $B\to K^{(*)}\ell\ell$ decays. The LHCb experiment provided through the years several measurements which seem to point to a coherent pattern of deviations \cite{Aaij:2014ora,Aaij:2017vbb,Aaij:2015oid,Aaij:2020nrf,Aaij:2019wad}. The consequences of these measurements are analysed in global fits, where different NP scenarios are studied \cite{Alguero:2019ptt,Aebischer:2019mlg,Ciuchini:2019usw, Hurth:2020rzx}. We chose to constrain the NP Wilson coefficients with their output.\\

Constraints from FCNCs also come from two-body leptonic decays, including LFV modes. The decay rate of a generic meson $P_{ij}= d^j \bar{d}^i$ into a lepton pair $\bar{\ell}_\alpha \ell_\beta$ generated by \eq{eq:lagrangian_bsll} reads
\begin{equation}
\begin{aligned}
\label{eq:Ptoll}
\mathcal{B}(P_{ij}\to\ell^-_\alpha \ell^+_\beta)=&\frac{\tau_P}{64\pi^3}\frac{\alpha_\text{EM}^2G_F^2}{m_P^3}|V_{t d_j} V^*_{t d_i}|^2 f_P^2 \, \lambda^{1/2}(m_P^2,m_\alpha^2,m_\beta^2) \times \\
\times&\left\{[m_P^2-(m_{\ell_\alpha}-m_{\ell_\beta})^2]\left|(m_{\ell_\alpha}+m_{\ell_\beta})\,\mathcal{C}^{ij\alpha\beta}_{10}\right|^2\right. \\
&+\left.[m_P^2-(m_{\ell_\alpha}+m_{\ell_\beta})^2]\left|(m_{\ell_\alpha}-m_{\ell_\beta})\,\mathcal{C}^{ij\alpha\beta}_{9}\right|^2\right\} \,.
\end{aligned}
\end{equation}
The full list of modes, together with their experimental bounds, is displayed in Table~\ref{tab:LFV_FCNC}. \\
Beyond the tree-level matching, $d_j\to d_i\ell_\alpha \ell_\beta$ receive contributions from box diagram. We checked using ref.~\cite{Bauer:2015knc,Gherardi:2020qhc} that these contributions are numerically irrelevant for our analysis.

\begin{table}
\begin{center}
\renewcommand{\arraystretch}{1.2} 
\begin{tabular}{c c }
\toprule
Observable & Upper limit  \\
\midrule
$\bar B_d\to\tau^-\mu^+$ & $2.2\cdot 10^{-5}$ \cite{Aubert:2008cu}  \\ 
$\bar B_d\to\tau^- e^+$ & $2.8 \cdot 10^{-5}$ \cite{Aubert:2008cu,Aaij:2019okb}  \\ 
$\bar B_d\to\mu^\pm e^\mp$ & $3.7 \cdot 10^{-9}$ \cite{Aaij:2017cza} \\ 
$\bar B_s\to\mu^\pm e^\mp$ & $1.4 \cdot 10^{-9}$ \cite{Aaij:2017cza} \\ 
$K_L\to\mu^\pm e^\mp$ & $4.7\cdot 10^{-12}$ \cite{Ambrose:1998us}  \\ 
$\bar B_s\to \tau^\pm\mu^\mp$ & $4.2 \cdot 10^{-5}$ \cite{Aaij:2019okb}\\
$\bar B_s\to \tau^+\tau^-$ & $6.8 \cdot 10^{-3}$ \cite{Aaij:2017xqt}\\
\toprule
\end{tabular}
\caption{Experimental measurements of semileptonic LFV decays.}
\label{tab:LFV_FCNC}
\end{center}
\end{table}

\subsection{$d_j\to d_i\nu_\alpha \nu_\beta$}
\subsubsection{$B\to K^{(*)}\nu\bar\nu$}

In $b\to s\nu\bar\nu$ transitions, the only new-physics contribution comes from left-handed operators. The relevant Lagrangian is 
\be
\begin{aligned}
\mathcal{L}(b\to s\nu\bar\nu)= +\frac{4G_F}{\sqrt{2}}\frac{\alpha_\text{EM}}{4\pi}V_{tb}V^*_{ts}\left(\mathcal{C}_{BK}^\text{SM}\delta_{\alpha\beta}+\mathcal{C}_{\nu}^{23\alpha\beta}\right)(\bar{s}\gamma^\mu P_L b)(\bar{\nu}^{\alpha}\gamma_\mu (1-\gamma_5) \nu^{\beta})\,,
 \end{aligned}
\label{eq:lagrangian_bsnunu}
\ee
where $\mathcal{C}_{BK}^\text{SM}\approx -6.35$ \cite{Buras:2014fpa}. The remaining NP Wilson coefficients are generically given by
\begin{align}
\mathcal{C}_{\nu}^{23\alpha\beta} = & \,+\frac{v^2}{M^2}\frac{\pi}{\alpha_\text{EM}|V_{tb}V^*_{ts}|}\left([\mathcal{C}_{lq}^{(1)}]^{23\alpha\beta}-[\mathcal{C}_{lq}^{(3)}]^{23\alpha\beta}\right) \,.
\label{eq:matching_BKnunu}
\end{align}
As it can be seen from \eq{eq:lagrangian_bsnunu}, final states with different neutrino species do not interfere with the SM contribution and are heavily suppressed. We neglect them in the following.
The branching ratio for $B\to K^{*}\bar\nu\nu$ decays can be expressed as:
\begin{equation}
\mathcal{B}(B\to K^{*}\bar\nu\nu) = \mathcal{B}(B\to K^{*}\bar\nu\nu)|_\text{SM} \times \Bigg[ \frac{1}{3}\,\sum_{\alpha=1}^3\bigg| 1 +\frac{\mathcal{C}_\nu^{23\alpha\alpha}}{C_{BK}^\text{SM}}\bigg|^2+\sum_{\alpha\neq\beta}\bigg|\frac{\mathcal{C}_\nu^{23\alpha\beta}}{C_{BK}^\text{SM}}\bigg|^2\Bigg]\,.
\label{eq:BtoKnunu}
\end{equation}
The SM expectation for the $B\to K^{*}\bar\nu\nu$ branching ratios are calculated in ref.~\cite{Buras:2014fpa} and read:
\begin{align}
\mathcal{B}(B\to K^{+}\bar\nu\nu)\vert_\text{SM} =&  (3.98\pm0.43\pm0.19)\times 10^{-6} \,,\\
\mathcal{B}(B\to K^{*0}\bar\nu\nu)\vert_\text{SM} =&  (9.19\pm0.86\pm0.50)\times 10^{-6}\,.
\end{align}
The most updated experimental results for these modes read
\begin{align}
\mathcal{B}(B\to K^{+}\bar\nu\nu) <& \,1.6\times 10^{-5} \,\text{\cite{Grygier:2017tzo}}\,, \\
\mathcal{B}(B\to K^{*0}\bar\nu\nu) <& \, 2.7\times 10^{-5} \, \text{\cite{Grygier:2017tzo}} \,,\\
\mathcal{B}(B\to K^{*+}\bar\nu\nu) <&\, 4.0\times 10^{-5}\, \text{\cite{Lees:2013kla}}\,,
\end{align}
where all three results are obtained at the 90\% C.L. . We can then  combine the theoretical predictions with the experimental upper limits to obtain \cite{Grygier:2017tzo}:
\begin{equation}
\mathcal{R}_K = \frac{\mathcal{B}(B\to K^{+}\bar\nu\nu)}{\mathcal{B}(B\to K^{+}\bar\nu\nu)|_\text{SM}}< 3.9 \quad\text{and}\quad \mathcal{R}_{K^*} = \frac{\mathcal{B}(B\to K^{*0}\bar\nu\nu)}{\mathcal{B}(B\to K^{*0}\bar\nu\nu)|_\text{SM}}< 2.7.
\label{eq:limitsBKnunu}
\end{equation}

\subsubsection{$K ^+\to \pi^+ \nu \bar\nu$}

The short-distance dominated decays $K\to \pi \nu \bar{\nu}$ serve as very clean modes to look for BSM effects. The effective Hamiltonian can be written as~\cite{Buchalla:1993wq}
\begin{align}
\mathcal{L} (s\to d\nu\bar\nu)= -& \frac{2 G_F}{\sqrt{2}} \frac{\alpha_{\rm{EM}}}{\pi \sin^2\theta_W} y_\nu \big[ (1+ C_{\nu}^{21\alpha\beta}) (\bar{s}\gamma^\mu P_L d)(\bar{\nu}^{\alpha}\gamma_\mu P_L \nu^{\beta})\big]\,,
\end{align}
where
\bea
\label{eq:ynudef}
y_\nu \equiv \sum \limits_{i=c,t} V_{is}^* V_{id} X(m_i^2/m_W^2) \simeq -(8+2i) \times 10^{-4}\,.
\eea
The function $X(x)$ is given, at leading order, by
\bea
 X_0(x) = \frac{x}{8} \left[ -\frac{2+x}{1-x} + \frac{3 x -6}{(1-x)^2} \ln x \right]\,,
\eea
and the new-physics coefficient is parametrized as
\bea
C_\nu^{21\alpha \beta} =  \, - \frac{v^2}{M^2}\frac{\pi \sin^2\theta_W}{\alpha_\text{EM}\, y_\nu }\left([\mathcal{C}_{lq}^{(1)}]^{21\alpha\beta}-[\mathcal{C}_{lq}^{(3)}]^{21\alpha\beta}\right) \,.
\label{eq:WCKtopinunu}
\eea
The recent analysis of NA62 provides the first measurement of this decay~\cite{ICHEP_Kpinunu},
\begin{equation}
\mathcal{B}(K^+\to\pi^+\nu\bar\nu) = (11^{+4.0}_{-3.5}\pm 0.3)\times 10^{-11}\,,
\end{equation}
which compared to the SM prediction in ref.~\cite{Buras:2015yca} yields:
\begin{equation}
\label{eq:RKpinunu}
\mathcal{R}_K  = \frac{\mathcal{B}(K^+\to\pi^+\nu\bar\nu)_\text{exp}}{\mathcal{B}(K^+\to\pi^+\nu\bar\nu)_\text{SM}}= 1.34\pm 0.47 \,.
\end{equation}

\subsection{$u^j\to u^i \ell^\alpha \ell^\beta$}

The general dimension-six effective Lagrangian for $u^j \to u^i \ell^\alpha \ell^\beta$ FCNC transition can be written as
\begin{equation}
\label{eq:LagUpFCNC}
{\cal L}_\text{eff}\, =\, \frac{ 4 G_F }{\sqrt{2}} \frac{\alpha_\text{EM}}{4\pi} \sum_{k} \xi_k^{i j \alpha \beta}\;  Q_{k}^{i j \alpha \beta},
\end{equation}
with the four-fermion operators:
\begin{equation}
\label{eq:def-operators}
\begin{array}{ll}
Q_{7}^{i j \alpha \beta}=\left(\bar{u}^i \sigma_{\mu\nu} P_{R} u^j\right)F^{\mu\nu} \delta_{\alpha\beta}, & Q_{7}^{\prime\,i j \alpha \beta}=\left(\bar{u}^i \sigma_{\mu\nu} P_{L} u^j\right) F^{\mu\nu}\delta_{\alpha\beta}\,, \\[1ex]
Q_{9}^{i j \alpha \beta}=\left(\bar{u}^i \gamma_{\mu} P_{L} u^j\right)\left(\bar{\ell}^\beta \gamma^{\mu} \ell^\alpha\right), & Q_{9}^{\prime\,i j \alpha \beta}=\left(\bar{u}^i \gamma_{\mu} P_{R} u^j\right)\left(\bar{\ell}^\beta \gamma^{\mu} \ell^\alpha\right)\,, \\[1ex]
Q_{10}^{i j \alpha \beta}=\left(\bar{u}^i \gamma_{\mu} P_{L} u^j\right)\left(\bar{\ell}^\beta \gamma^{\mu} \gamma_{5} \ell^\alpha\right), & Q_{10}^{\prime\,i j \alpha \beta}=\left(\bar{u}^i \gamma_{\mu} P_{R} u^j\right)\left(\bar{\ell}^\beta \gamma^{\mu} \gamma_{5} \ell^\alpha\right)\,, \\[1ex]
Q_{S}^{i j \alpha \beta}=\left(\bar{u}^i P_{R} u^j\right)(\bar{\ell}^\beta \ell^\alpha), & Q_{S}^{\prime\,i j \alpha \beta}=\left(\bar{u}^i P_{L} u^j\right)(\bar{\ell}^\beta \ell^\alpha)\,, \\[1ex]
Q_{P}^{i j \alpha \beta}=\left(\bar{u}^i P_{R} u^j\right)\left(\bar{\ell}^\beta \gamma_{5} \ell^\alpha\right), & Q_{P}^{\prime\,i j \alpha \beta}=\left(\bar{u}^i P_{L} u^j\right)\left(\bar{\ell}^\beta \gamma_{5} \ell^\alpha\right)\,, \\[1ex]
Q_{T}^{i j \alpha \beta}=\left(\bar{u}^i \sigma^{\mu \nu} u^j\right)\left(\bar{\ell}^\beta \sigma_{\mu \nu} \ell^\alpha\right), & Q_{T 5}^{i j \alpha \beta}=\left(\bar{u}^i \sigma^{\mu \nu} u^j\right)\left(\bar{\ell}^\beta \sigma_{\mu \nu} \gamma_{5} \ell^\alpha\right)\,.
\end{array}
\end{equation}

In the SM the FCNC decays of $D$ mesons are highly GIM-suppressed and are dominated by the resonance contributions. For example, for the decay $D^+\to \pi^+ \ell\ell$, the short distance contribution to the Wilson coefficients $\xi_{9}$ estimates the branching fraction about four orders of magnitude smaller than the current experimental bound. Hence, in order to constrain the NP parameter space, we neglect the SM contribution to the Wilson coefficients. The matching with the Lagrangian in Eq.~\eqref{eq:S1S3L} gives
\begin{align}
\mathbb{\xi}_{9}^{i j \alpha \beta} &=-\xi_{10}^{i j \alpha \beta}=+\frac{v^{2}}{2M^{2}} \frac{\pi}{\alpha_{\mathrm{EM}}} \left(|g_1|^2 (V^* S_{QL})^{j\beta}\; (V S_{QL}^*)^{i\alpha}+|g_3|^2 (V^* \tilde S_{QL})^{j\beta}\; (V \tilde S_{QL}^*)^{i\alpha}\right)\,, \\
\xi_{9}^{\prime\, i j \alpha \beta} &=+\xi_{10}^{\prime\, i j \alpha \beta}=+\frac{v^{2}}{2M^{2}} \frac{\pi}{\alpha_{\mathrm{EM}}} |g_R|^2 S_{UE}^{j\beta}\; S_{UE}^{*\,i\alpha}\,, \\
\xi_{S}^{i j \alpha \beta} &=+\xi_{P}^{i j \alpha \beta}=-\frac{v^{2}}{2M^{2}} \frac{\pi}{\alpha_{\mathrm{EM}}}\; g_1^* g_R (V S_{QL}^*)^{i\alpha}S_{UE}^{j\beta}\,, \\
\xi_{S}^{\prime\, i j \alpha \beta} &=-\xi_{P}^{\prime\,i j \alpha \beta}=-\frac{v^{2}}{2M^{2}} \frac{\pi}{\alpha_{\mathrm{EM}}}\; g_1 g_R^* (V^* S_{QL})^{j\beta}\; S_{UE}^{*\,i\alpha}\,, \\
\xi_{T}^{i j \alpha \beta} &= -\frac{1}{8} \left( \xi_{S}^{ i j \alpha \beta} +\xi_{S}^{\prime\, i j \alpha \beta}  \right)\,, \\
\xi_{T5}^{ i j \alpha \beta} &= -\frac{1}{8} \left( \xi_{S}^{ i j \alpha \beta} -\xi_{S}^{\prime\, i j \alpha \beta}  \right)\,.
\end{align}

We note that the matching above is done at the NP scale $M$. Assuming $M \sim 2\,$TeV and using the RG equations given in Eq.~\eqref{eq:running} the couplings at the relevant scale for $D$ decays $\mu=\overline{m}_c(m_c)$ are
\bea
\xi_S(\overline{m}_c) = 2.42\, \xi_S (M)~~{\rm and}~~ \xi_T(\overline{m}_c) = 0.74\, \xi_T (M)\,.
\eea

\subsubsection{$D_0\to \ell \ell$}

The branching fraction for these decays is given by
\begin{align}
\mathcal{B}&\left(P_{i j} \rightarrow \ell_{\alpha}^{+} \ell_{\beta}^{-}\right)= \frac{\tau_{P}}{64 \pi^{3}} \frac{\alpha_{\mathrm{EM}}^{2} G_{F}^{2}}{m_{P}^{3}} f_{P}^{2} \lambda^{1 / 2}\left(m_{P}^{2}, m_{\alpha}^{2}, m_{\beta}^{2}\right) \times \nn \\
&\left\{\left[m_{P}^{2}-\left(m_{\ell_{\alpha}}-m_{\ell_{\beta}}\right)^{2}\right]\left|\left(m_{\ell_{\alpha}}+m_{\ell_{\beta}}\right)\left(\xi_{10}^{i j \alpha \beta}-\xi_{10}^{\prime \,i j \alpha \beta}\right)+\frac{m_{P}^{2}}{m_{i}+m_{j}} \left(\xi_{P}^{i j \alpha \beta}-\xi_{P}^{\prime\,i j \alpha \beta}\right)\right|^{2}\right. \nn \\
&\left.+\left[m_{P}^{2}-\left(m_{\ell_{\alpha}}+m_{\ell_{\beta}}\right)^{2}\right]\left|\left(m_{\ell_{\alpha}}-m_{\ell_{\beta}}\right)\left(\xi_{9}^{i j \alpha \beta}-\xi_{9}^{\prime\, i j \alpha \beta}\right)+\frac{m_{P}^{2}}{m_{i}+m_{j}} \left(\xi_{S}^{i j \alpha \beta}-\xi_{S}^{\prime\,i j \alpha \beta}\right)\right|^{2}\right\}\,.
\end{align}
The current upper limits at 90\% C.L.  are 
\begin{eqnarray}
\mathcal{B} (D^0\to e^+e^-) <& 7.9 \times 10^{-8} \text{~\cite{Petric:2010yt}}\,, \nn \\
\mathcal{B} (D^0\to \mu^+\mu^-) <& 6.2 \times 10^{-9}\text{~\cite{Aaij:2013cza}}\,,\\
\mathcal{B} (D^0\to e^\mp \mu^\pm) <& 1.3 \times 10^{-8}\text{~\cite{Aaij:2015qmj}}\,. \nn
\end{eqnarray}

\subsection{$\Delta F =2$}
\subsubsection{$B_q-\bar{B}_q$ mixing}
\label{app:Bmixing}

The effective Hamiltonian describing $B^0_q-\bar{B}^0_q$ mixing can be parametrized as
\begin{equation}
\mathcal{H}_{\rm{eff}}^{\Delta B=2} = \left[\frac{G_F^2}{16\pi^2}m_W^2(V_{tb}^*V_{tq})^2+\mathcal{C}_{qq}\right]({\bar{q}}\gamma_\mu P_L b)({\bar{q}}\gamma^\mu P_L b)+{\rm{h.c.}}
\label{eq:lagr_Bmix}
\end{equation} 
The hadronic matrix element of the single $\Delta B=2$ operator is
\begin{equation}
\langle\bar{B}^0_q|({\bar{q}}\gamma_\mu P_L b)({\bar{q}}\gamma^\mu P_L b)(\mu)|B^0_q\rangle = \frac{1}{3}m_{B_q}f_{B_q}^2B_q^{VLL}(\mu)\,,
\end{equation}
where $f_{B_q}$ is the meson decay constant and $B_q^{VLL}$ the bag parameter. The real and imaginary parts of the matrix element $\mathcal{M}(B^0_q\to\bar{B}^0_q)\equiv \mathcal{M}_{12}(B_q)$ are related to the meson mass difference and the mixing angle as 
\begin{equation}
\Delta M_q = 2 |\mathcal{M}_{12}(B_q)|\,, \qquad \text{and} \qquad \phi_b =\text{Arg} [\mathcal{M}_{12}(B_q)]\,.
\end{equation}
In the SM, one finds
\begin{equation}
\mathcal{M}_{12}(B_q)|_\text{SM}=\frac{\GF^2 m_W^2 m_{B_q}}{12\pi^2}(V_{tb}V_{tq}^*)^2 f_{B_q}^2\hat{\eta}_B S_0(x_t) B_q^{VLL} \,,
\end{equation}
where $S_0(x_t)\approx 2.36853$ is the Inami-Lim function defined in ref.~\cite{Inami:1980fz} and $\hat{\eta}_B\approx 0.842$ \cite{Buras:2001ra} encodes the QCD running from $\mu=m_t$ to $\mu=m_b$. In the presence of NP, one finds 
\begin{equation}
\mathcal{M}_{12}(B_q) = \mathcal{M}_{12}(B_q)|_\text{SM}\left[1+\frac{4\pi^2\mathcal{C}_{qq}}{G_F^2 m_W^2(V_{tb}^*V_{tq})^2 S_0(x_t)}\right] \,.
\end{equation}
In the presence of $S_1$ and $S_3$ leptoquarks, the new-physics coefficient takes the form~\cite{Crivellin:2019dwb}
\begin{equation}
\begin{aligned}
\mathcal C_{qq} = \frac{1} {128 \pi^2 M^2}\eta_\text{LQ} \bigg[&|g_1|^4 S_{QL}^{*q\alpha}S_{QL}^{3\beta}S_{QL}^{*q\beta}S_{QL}^{3\alpha}+5|g_3|^4 \tilde S_{QL}^{*q\alpha} \tilde S_{QL}^{3\beta} \tilde S_{QL}^{*q\beta} \tilde S_{QL}^{3\alpha} \\
+&2|g_1|^2|g_3|^2 S_{QL}^{*q\alpha} S_{QL}^{3\beta} \tilde S_{QL}^{*q\beta} \tilde S_{QL}^{3\alpha}\bigg]\,,
\end{aligned}
\end{equation}
where the factor $\eta_\text{LQ}$ takes into the account the running from the leptoquark scale to the electroweak scale. With the package \texttt{WCxf} \cite{Aebischer:2018bkb} one finds $\eta_\text{LQ}\approx 1$. Since we are considering real coefficients, only the real part of the amplitude receives NP corrections. 

For our numerical analysis, we use the results in ref.~\cite{DiLuzio:2019jyq}, where the weighted average for the matrix elements in refs.~\cite{Aoki:2019cca,Kirk:2017juj,King:2019lal,Dowdall:2019bea,Boyle:2018knm} is used to yield
\begin{equation}
\begin{aligned}
    \Delta M_d^\text{average} =&\, (1.05^{+0.04}_{-0.07})\, \Delta M_d^\text{exp}\,, \\
    \Delta M_s^\text{average} =&\, (1.04^{+0.04}_{-0.07})\, \Delta M_s^\text{exp}\,. 
\end{aligned}
\end{equation}

\subsubsection{$K^0-\bar{K^0}$ mixing}

The  $\Delta S=2$ effective Hamiltonian is given by
\begin{align}
\mathcal{H}_{\rm{eff}}^{\Delta S=2} = \left\{ \frac{G_F^2 m_W^2}{4\pi^2} C_{\rm SM} + [\mathcal{C}_{qq}]^{1212} \right\} (\bar{s} \gamma^\mu P_L d)(\bar{s} \gamma_\mu P_L d)+{\rm{h.c.}}
\end{align}

The contribution to the off-diagonal matrix element is defined as $M_{12}=\displaystyle\frac{\langle K^0|\mathcal{H}_{\Delta S=2}|\bar{K^0}\rangle}{2m_K}$, with
\bea
\langle K^0|(\bar{d}_{L,R} \gamma^\mu s_{L,R})^2|\bar{K^0}\rangle = {4\over 3} f_K^2 \hat{B}_K m_K^2\,,
\eea
where $f_K$ is the kaon decay constant and $\hat{B}_K$ the reduced bag parameter.

The SM contribution to the Wilson coefficient reads 
\begin{align}
C_{\rm SM}= {\kappa_{c}^2} \eta_{cc} S_0(x_c)+ {\kappa_t^2} \eta_{tt} S_0(x_t)+ 2\kappa_c \kappa_t \eta_{ct} S_0(x_c,x_t) \, ,
\end{align}
where we used the short-hand notation for CKM factors $\kappa_i= V_{is}^* V_{id}$, $S_0(x_i)$ are the Inami-Lim functions and the $\eta_i$ factors account for QCD effects (see e.g. ref.~\cite{Buras:1998raa} for details). The Wilson coefficient $[\mathcal{C}_{qq}]^{1212}$ gets contributions from $S_1$ and $S_3$ at one-loop level~\cite{Bobeth:2017ecx}:
\begin{equation}
\begin{aligned}
\left[\mathcal C_{qq}\right]^{1212} = \frac{1} {128 \pi^2 M^2}\eta_\text{LQ} \bigg[&|g_1|^4 S_{QL}^{*1\alpha}S_{QL}^{2\beta}S_{QL}^{*1\beta}S_{QL}^{2\alpha}+5|g_3|^4 \tilde S_{QL}^{*1\alpha} \tilde S_{QL}^{2\beta} \tilde S_{QL}^{*1\beta} \tilde S_{QL}^{2\alpha} \\
+&2|g_1|^2|g_3|^2 S_{QL}^{*1\alpha} S_{QL}^{2\beta} \tilde S_{QL}^{*1\beta} \tilde S_{QL}^{2\alpha}\bigg]\,.
\end{aligned}
\end{equation}

The meson mass difference $\Delta m_K$ can be obtained from
\begin{eqnarray}
\label{eq:obs}
\Delta m_K \approx 2\, \Re\, M_{12}\,,
\end{eqnarray}
and is to be compared with the experimental value~\cite{Tanabashi:2018oca}
\begin{equation}
\Delta m_K = (3.484\pm 0.006)\times 10^{-15}\,{\rm GeV}\,.
\end{equation}

\subsubsection{$D^0-\bar{D}^0$ mixing}

The relevant effective Hamiltonian is~\cite{Golowich:2009ii}
\begin{equation}
\mathcal{H}_\text{eff}^{\Delta C=2}= \sum_{i=1}^4 \mathcal{C}_i(\mu) \mathcal{Q}_i(\mu)\,,
\label{eq:lagr_Dmixing}
\end{equation}
where
\begin{equation}
\begin{aligned}
\mathcal{Q}_1 &= (\bar{u}\gamma_\mu P_L c)(\bar{u}\gamma^\mu P_L c)\,,  & \mathcal{Q}_2 &= (\bar{u}\gamma_\mu P_L c)(\bar{u}\gamma^\mu P_R c)\,, \\
\mathcal{Q}_3 &= (\bar{u} P_R c)(\bar{u} P_L c)\,, & \mathcal{Q}_6 &= (\bar{u}\gamma_\mu P_R c)(\bar{u}\gamma^\mu P_R c)\,.
\end{aligned}
\end{equation}
Integrating out the leptoquarks and matching yields the following values for the coefficients:
\begin{equation}
    \begin{aligned}
    \mathcal{C}_1(M) =&\,  \frac{1}{128\pi^2 M^2}\bigg[|g_1|^2 S_{QL}^{2i} S_{QL}^{*1i} S_{QL}^{2j} S_{QL}^{*1j}+5 |g_3|^4  \tilde S_{QL}^{2i} \tilde S_{QL}^{*1i} \tilde S_{QL}^{2j} \tilde S_{QL}^{*1j}+2 |g_1|^2 |g_3|^2 S_{QL}^{2i} S_{QL}^{*1i} \tilde S_{QL}^{2j} \tilde S_{QL}^{*1j}\bigg] \,, \\
    \mathcal{C}_2(M) =&\,  -\frac{1}{128\pi^2 M^2} 2 |g_1|^2 |g_R|^2 S_{QL}^{2i} S_{QL}^{*1i} S_{UE}^{*1j} S_{UE}^{2j}\,,\\
    \mathcal{C}_3(M) =&\,  0 \,,  \\
    \mathcal{C}_6(M) =&\,  \frac{1}{128\pi^2 M^2} |g_R|^4S_{UE}^{2i}S_{UE}^{*1i} S_{UE}^{2j} S_{UE}^{*1j}\,.
    \end{aligned}
\end{equation}
RG running down to $\mu = 3\,\text{GeV}$ is done with the package \texttt{WCxf}~\cite{Aebischer:2017ugx}. Together with the values of the matrix elements listed in \Table{tab:Dmixing}, our expression for $M_{12}$ reads:
\begin{equation}
\begin{aligned}
    M_{12} = \frac{1}{2 m_D}\bigg[&\mathcal{C}_1(M) \eta_\text{high}^{\text{LL}} \eta_\text{EW}^{\text{LL}}\langle \mathcal{Q}_1\rangle + \mathcal{C}_6(M)\eta_\text{high}^{\text{RR}} \eta_\text{EW}^{\text{RR}}\langle \mathcal{Q}_6\rangle+\mathcal{C}_2(M) \eta_\text{high}^{\text{LR}}\left(\eta_\text{EW}^{\text{LR}}\langle\mathcal{Q}_2\rangle+\eta_\text{EW}^{\text{S}_\text{LR}}\langle\mathcal{Q}_3\rangle\right)\bigg]\,,
    \end{aligned}
\end{equation}
where
\begin{equation}
    \begin{aligned}
    \eta_\text{high}^{\text{LL}} =&\,  0.98\,, &
    \eta_\text{high}^{\text{RR}} =&\,  0.91\,, &
    \eta_\text{high}^{\text{LR}}=&\,1\,,\\ \eta_\text{EW}^{\text{LL}}=\, \eta_\text{EW}^{\text{RR}} =&\, 0.79\,,  & \eta_\text{EW}^{\text{LR}}=&\, 0.91\,, & \eta_\text{EW}^{\text{S}_\text{LR}}=& -1\,.
    \end{aligned}
\end{equation}

With real new-physics couplings, deviations from the SM on charm mixing are sensitive to
\begin{equation}
x_D = \frac{2 |M_{12}|}{\Gamma_D^0}\,.
\end{equation}
The HFLAV collaboration determined this quantity from a global fit and obtained $x_D^\text{exp}= (4.1^{+1.4}_{-1.5})\times 10^{-3}$ ~\cite{Amhis:2019ckw}. 

\begin{table}
\begin{center}
\renewcommand{\arraystretch}{1.2} 
\begin{tabular}{c c }
\toprule
Matrix element & Value  \\
\midrule
$\langle \mathcal{Q}_1\rangle$ & 0.0805(55)\,   \\ 
$\langle \mathcal{Q}_2\rangle$&  -0.2070(142)  \\ 
$\langle \mathcal{Q}_6\rangle$&   0.0805(55) \\ 
$\langle \mathcal{Q}_3\rangle$&   0.2747(129) \\ 
\toprule
\end{tabular}
\caption{Values for the hadronic matrix elements in $\Delta C=2$ processes at the scale $\mu=3 \, \text{GeV}$ from ref.~\cite{Bazavov:2017weg}. The matrix elements are expressed in $\text{GeV}^4$.}
\label{tab:Dmixing}
\end{center}
\end{table}
\subsection{$Z\to \bar{\ell}^\beta \ell^\alpha$ and $Z\to \bar{\nu}^\beta \nu^\alpha$}
\label{app:Ztoll}

The importance of corrections to $Z$ couplings has been pointed out in refs.~\cite{Feruglio:2016gvd,Feruglio:2017rjo}. They are described through the following Lagrangian:
\begin{equation}
\mathcal{L}_\text{eff}^Z = \frac{g}{\cos\theta_w\sin\theta_w}\sum_{f,\alpha,\beta}\bar{f}_\beta\gamma_\mu[\mathcal{G}_{f_L}^{\alpha\beta} P_L+\mathcal{G}_{f_R}^{\alpha\beta}P_R]f_\alpha Z^\mu\,,
\end{equation}
where $f_i$ are the SM fermions and
\begin{align}
\mathcal{G}_{f_{L(R)}}^{\alpha\beta} = \delta^{\alpha\beta}g_{f_{L(R)}}+\delta g^{\alpha\beta}_{f_{L(R)}}\,.
\end{align}
The couplings $g_{f_{L(R)}}$ are the SM ones, defined as: $g_{f_L} = I^3_{f_L}-Q_{f}\sw$ and $g_{f_R} = -Q_{f}\sw$, while $\delta g^{\alpha\beta}_{f_{L(R)}}$ are the corresponding NP penguin corrections. Taking the computations in ref.~\cite{Arnan:2019olv} and adapting them to our scenario, we find (denoting $\xt=m_t^2/M^2$ and $\xz=m_Z^2/M^2$)
\begin{align}
\label{eq:znunu_expression}
\delta g^{\alpha\beta}_{\nu} &= \frac{N_c}{8\pi^2}  |g_3|^2(V^* \tilde S_{QL})^{3\alpha} (V \tilde S_{QL}^*)^{3\beta}  \left[(g_{t_L}-g_{t_R})\frac{\xt(\xt-1-\log(\xt))}{(1-\xt)^2}+\frac{\xz}{12}F^L (\xt)\right] \nonumber\\
&+ \frac{\xz N_c}{24\pi^2} |g_3|^2\sum_{k=u,c} (V^* \tilde S_{QL})^{k\alpha}(V \tilde S_{QL}^*)^{k\beta}\left[g_{u^k_L}\left(\log(\xz)-i\pi-\frac{1}{6}\right)+\frac{g_{\nu_L}}{6}\right] \nonumber\\
&+ \frac{\xz N_c}{48\pi^2} \sum_{k=d,s,b} 
\left( |g_3|^2 \tilde S_{QL}^{k\alpha} \tilde S_{QL}^{*k\beta}+ |g_1|^2 S_{QL}^{k\alpha} S_{QL}^{*k\beta} \right) \left[g_{d^k_L}\left(\log(\xz)-i\pi-\frac{1}{6}\right)+\frac{g_{\nu_L}}{6}\right] \,,
\end{align}
for the neutrino couplings and  
\begin{align}
\label{eq:zll_expression}
\delta g^{\alpha\beta}_{\ell_L} &= \frac{N_c}{16\pi^2}\left( |g_3|^2(V^* \tilde S_{QL})^{3\alpha} (V  \tilde S_{QL}^*)^{3\beta}+ |g_1|^2(V^* S_{QL})^{3\alpha} (V S_{QL}^*)^{3\beta} \right) \times \nn \\ &\qquad\left[(g_{t_L}-g_{t_R})\frac{\xt(\xt-1-\log(\xt))}{(1-\xt)^2}+\frac{\xz}{12}F^L (\xt)\right] \nonumber\\
&+ \frac{\xz N_c}{48\pi^2} \sum_{k=u,c} \left(|g_3|^2 (V^* \tilde S_{QL})^{k\alpha}(V \tilde S_{QL}^*)^{k\beta} + |g_1|^2 (V^* S_{QL})^{k\alpha} (V S_{QL}^*)^{k\beta} \right) \times \nonumber\\
& \qquad \quad \left[g_{u^k_L}\left(\log(\xz)-i\pi-\frac{1}{6}\right)+\frac{g_{\ell_L}}{6}\right] \nonumber\\
&+ \frac{\xz N_c}{24\pi^2} |g_3|^2\sum_{k=d,s,b} (\tilde S_{QL})^{k\alpha}(\tilde S_{QL}^*)^{k\beta}\left[g_{d^k_L}\left(\log(\xz)-i\pi-\frac{1}{6}\right)+\frac{g_{\ell_L}}{6}\right] \,,
\end{align}
for the left-handed charged lepton couplings. A similar expression holds for the right-handed ones with trivial substitutions. The function $F^L(\xt)$ reads
\begin{equation}
\begin{aligned}
F^L(\xt) =& +g_{t_L}\frac{(1-\xt)(5 \xt^2-7\xt+8)+2(\xt^3+2)\log(\xt)}{(1-\xt)^4} \\
&+g_{t_R}\frac{(1-\xt)(\xt^2-5\xt-2)-6 \xt\log(\xt)}{(1-\xt)^4} \\
&-g_{\ell_L}\frac{(1-\xt)(-11\xt^2+7\xt-2)-6\xt^3\log(\xt)}{3(1-\xt)^4}\,.
\end{aligned}
\label{eq:Zllloop}
\end{equation}
The expression for $\delta g_{\ell_R}^{\alpha\beta}$ can be obtained from \eqs{eq:zll_expression}{eq:Zllloop} by replacing $L\to R$ and $\left(|g_3|^2 (V^* \tilde S_{QL})^{3\alpha} (V \tilde S_{QL}^*)^{3\beta}+ |g_1|^2(V^* S_{QL})^{3\alpha} (V S_{QL}^*)^{3\beta} \right)\to |g_R|^2S_{UE}^{3\alpha} S_{UE}^{*3\beta}$ for the first two lines in \eq{eq:zll_expression} only. The spurion associated with $S_3$ does not generate such contribution.

The allowed values for the quantities $g^{\alpha\beta}_{\ell_{L(R)}} $ and $g^{\alpha\beta}_{\nu}$ can be found in ref.~\cite{Falkowski:2019hvp}. The dominant contribution comes from the top-mediated contribution (singled out in the first line of both expressions), whose magnitude is ${\cal{O}}(10^{-4})$ for $M= 2 \, \text{TeV}$. The remaining contributions are suppressed by at least one order of magnitude.

\subsection{$W$ LFU}
\label{app:WLFU}
The corrections to $W$ coupling  have been first analysed in  refs.~\cite{Feruglio:2016gvd,Feruglio:2017rjo}.
The effective Lagrangian describing them reads:
\begin{equation}
    \mathcal{L} = -\frac{g}{\sqrt{2}}(\delta^{ij}+ \delta g_W^{ij})\,\bar{e}_i\gamma^\mu P_L \nu_j W_\mu + \text{h.c.}
    \label{eq:Wcorrections}
\end{equation}
Both $S_1$ and $S_3$ contribute via penguin diagrams to $\delta g_W^{ij}$. Their expressions are obtained in ref.~\cite{Arnan:2019olv} and read:
\begin{align}
    \left[g_W^{ij}\right]_{S_1} =\,& -\frac{N_c}{64 \pi^2} |g_1|^2 (V^* S_{QL})_{3i}(V  S^*_{QL})_{3j}\bigg[-\frac{\xt(\xt-1+(\xt-2)\log(\xt))}{(\xt-1)^2}+\frac{2}{9} x_W G_{S_1}(\xt)\bigg] \nonumber\\
    &-\,\frac{4}{9} x_W|g_1|^2\sum_{k=u,c}(V^* S_{QL})_{ki}(V S^*_{QL})_{kj} \left(-1-3\log(x_W)+3\pi i \right)\,, \label{eq:Wloop_S1}\\
    \left[g_W^{ij}\right]_{S_3} =\,& -\frac{N_c}{64 \pi^2} |g_3|^2(V^* \tilde S_{QL})_{3i}(V  \tilde S^*_{QL})_{3j}\bigg[+\frac{\xt(\xt-1+(\xt-2)\log(\xt))}{(\xt-1)^2}+\frac{2}{9} x_W G_{S_3}(\xt)\bigg] \nonumber\\
    &-\,\frac{4}{9} x_W |g_3|^2\sum_{k=u,c}(V^* \tilde S_{QL})_{ki}(V \tilde S^*_{QL})_{kj} \left(1-3\log(x_W)+3\pi i \right)\,,\label{eq:Wloop_S3}
\end{align}
where $x_i = m_i^2/M^2$ and 
\begin{align}
    G_{S_1}(x_t) =&\, \frac{6(\xt-1-\log\xt)}{(\xt-1)^2}\,, \\
    G_{S_3}(x_t) =&\, \frac{6[\xt(\xt^2+\xt-2)+1]\log\xt+\xt-[\xt(\xt(2\xt-23)+15)-10]}{(\xt-1)^4}\,.
\end{align}
The values allowed by experimental data for $\delta g_W^{ij}$ can be found in ref.~\cite{Falkowski:2019hvp} and read:
\begin{align}
    \delta g_W^{11}=& -0.0050\pm 0.0031\,, & \delta g_W^{22}=& -0.0140\pm0.0050\,, &  \delta g_W^{33}=& +0.0163\pm0.0060\,.
\end{align}
In a recent analysis of the ATLAS collaboration \cite{Aad:2020ayz},  the universality of lepton weak couplings has been measured  through the ratio 
\begin{equation}
    R(\tau/\mu) = \frac{\mathcal{B}(W\to\tau\bar\nu)}{\mathcal{B}(W\to\mu\bar\nu)}= 0.992\pm 0.013\,,
\end{equation}
while the SM expectation is $R(\tau/\mu)\approx 1$ even when taking into account phase space effects. We stress that this result is in agreement with the SM expectation, while previous results from the LEP experiments in ref.\cite{Schael:2013ita} used in ref.~\cite{Falkowski:2019hvp} showed an upward deviation with respect to the SM expectation of $2.7\,\sigma$. In the main text, we comment on the consequences of this result.

\subsection{LFV  lepton decays}
\subsubsection{$\ell^\beta\to\ell^\alpha \gamma$}

The radiative decay $\ell^\beta\to\ell^\alpha \gamma$ takes place predominantly through penguin diagrams involving the exchange of quarks and leptoquarks. The box contributions are suppressed and are not considered. The decay width (neglecting the light lepton mass) can be written as~\cite{Mandal:2019gff}
\begin{eqnarray}
\label{eq:muegamma}
{\Gamma}(\ell^\beta \to \ell^\alpha \gamma) &=& \frac{\alpha_{EM}}{4}\,m^3_{\ell^\beta}\; \big(|A_L^{\alpha\beta}|^2 +|A_R^{\alpha\beta}|^2 \big),
\end{eqnarray}
where
\begin{eqnarray}
\label{eq:AR}
A_L^{\alpha\beta} &=& \frac{3}{32 \pi^2} \frac{1}{M^2}
\; \Bigg\{ m_{\ell_i} \sum_{j=u,c,t} \!\!
\bigg( |g_3|^2 (V^* \tilde S_{QL})^{j\beta} (V \tilde S_{QL}^*)^{j\alpha} +  |g_1|^2 (V^*S_{QL})^{j\beta} (V S_{QL}^*)^{j\alpha}  \bigg)
\nn \\ 
&&\hskip 4cm\mbox{} \times 
\bigg[ -\frac{2}{3} F_{1}(x_j) + \frac{1}{3} F_{2}(x_j)\bigg] 
\nn \\ 
&&\hskip 1.2cm\mbox{} + m_{\ell_i} \sum_{j=d,s,b} \! 2|g_3|^2\, \tilde S_{QL}^{j\beta} \tilde S_{QL}^{*j\alpha} \,
\bigg[ \frac{1}{3} F_{1}(x_j) + \frac{4}{3} F_{2}(x_j)\bigg] 
\nn \\ 
&&\hskip 1.2cm\mbox{}- m_{q_j} \sum_{j=u,c,t}  g_1 g_R^* \,
(V^* S_{QL})^{j\beta}\, S_{UE}^{*j\alpha} \, \bigg[ -\frac{2}{3} F_{3}(x_j) + \frac{1}{3}
F_4 (x_j)\bigg]\Bigg\}\,, \\
A_R^{\alpha\beta} &=& \frac{3}{32 \pi^2} \frac{1}{M^2}
\; \Bigg\{ m_{\ell_i} \sum_{j=u,c,t} \!\!  |g_1|^2 S_{UE}^{j\beta} S_{UE}^{*j\alpha} 
\bigg[ -\frac{2}{3} F_{1}(x_j) + \frac{1}{3} F_{2}(x_j)\bigg] 
\nn \\ 
&&\hskip 1.2cm\mbox{}- m_{q_j} \sum_{j=u,c,t}  g_1^* g_R \,
S_{UE}^{j\beta}\, (V S_{QL}^*)^{j\alpha} \, \bigg[ -\frac{2}{3} F_{3}(x_j) + \frac{1}{3}
F_4 (x_j)\bigg]\Bigg\}\,.
\end{eqnarray}
With $x_j=m_{q_j}^2/ M^2$ the loop functions are defined as
\begin{equation}
\label{eq:loop_lgamma}
\begin{aligned}
F_1(x_j) =& 
\frac{1}{6 \,(1-x_j)^4} \, (2+3 \, x_j-6 \, x_j^2+x_j^3+6 \, x_j  \, \ln x_j) \;,  \\
F_2(x_j) =& \frac{1}{6 \, (1-x_j)^4} \, (1-6 \, x_j+3 \, x_j^2+2 \, x_j^3-6 \, x_j^2 \, \ln x_j) \; ,
\\
F_3(x_j) =&
\frac{1}{(1-x_j)^3} \, (-3+4 \, x_j-x_j^2-2 \,  \ln x_j) \; ,
 \\
F_4(x_j) =& \frac{1}{(1-x_j)^3} \, (1-x_j^2+2 \, x_j \,\ln x_j) \;.
\end{aligned}
\end{equation}

The most stringent upper limits on $\mu\to e\gamma$ and $\tau \to  \ell\gamma$ are provided by the MEG experiment~\cite{TheMEG:2016wtm} and BaBar~\cite{Aubert:2009ag}, respectively. The current 90\% C.L. limits are:
\begin{eqnarray}
\label{eq:datalgamma}
\mathcal{B}(\mu\to e\gamma)&<& 4.2\times 10^{-13}, \\
\mathcal{B}(\tau\to e\gamma)&<& 3.3\times 10^{-8},\\
\label{eq:datalgamma3}
\mathcal{B}(\tau\to \mu\gamma)&<& 4.4\times 10^{-8}.
\end{eqnarray}

\subsubsection{$\tau\to3\ell$}

Following the discussion in ref.\cite{Mandal:2019gff}, we have
\begin{equation}
    \begin{aligned}
    \mathcal{B}(\tau\to3\ell) = \frac{\alpha_\text{EM}m_\tau^5}{32\pi \Gamma_\tau}&\bigg\{|T_{1L}|^2+|T_{1R}|^2+(|T_{2L}|^2+|T_{2R}|^2)\left(\frac{16}{3}\log\frac{m_\tau}{m_\mu}-\frac{22}{3}\right) \\
    &-4 \Re[T_{1L}T^*_{2R}+T_{2L}T^*_{1R}]\bigg\}\,,
    \end{aligned}
\end{equation}
where we retain only the leading contributions from penguin diagrams. The loop functions read:
\begin{align}
    T_{1L} =& -\frac{3}{16\pi^2}\frac{1}{M^2}|g_1|^2 (V^* S_{QL})^{j3} (V S^*_{QL})^{j2} \left[-\frac{2}{3}\left(\frac{4}{9}+\frac{1}{3}\log x_j\right)+\frac{1}{54} \right]\,, \\
    T_{1R} =& -\frac{3}{16\pi^2}\frac{1}{M^2}|g_R|^2  S_{UE}^{j3} S_{UE}^{*j2} \left[-\frac{2}{3}\left(\frac{4}{9}+\frac{1}{3}\log x_j\right)+\frac{1}{54} \right]\,, \\
    T_{2L}=&-\frac{3}{16\pi^2}\frac{1}{M^2}\bigg\{-\frac{2}{3}\left[\frac{1}{6} S_{UE}^{j3}S_{UE}^{*j2}- \frac{m_{q_j}}{m_{\ell_j}}S_{UE}^{j3} (V S_{QL}^{*})^{j2}\left(\frac{3}{2}+\log x_j\right)\right] \nonumber\\
    &+\frac{1}{3}\left(\frac{1}{12}S_{UE}^{j3}S_{UE}^{*j2}-\frac{1}{2}S_{UE}^{j3}(V S_{QL}^{*})^{j2}\right)\bigg\}\,,\\
    T_{2R}=&-\frac{3}{16\pi^2}\frac{1}{M^2}\bigg\{-\frac{2}{3}\left[\frac{1}{6} (V^* S_{QL})^{j3} (V S^*_{QL})^{j2}- \frac{m_{q_j}}{m_{\ell_j}}(V^* S_{QL})^{j3} S_{UE}^{*j2}\left(\frac{3}{2}+\log x_j\right)\right]\nonumber \\
    &+\frac{1}{3}\left(\frac{1}{12}(V^* S_{QL})^{j3} (V S^*_{QL})^{j2}-\frac{1}{2}(V^* S_{QL})^{j3} S_{UE}^{*j2}\right)\bigg\}\,,
\end{align}
where $x_j = m_{q_j}^2/M^2$.
The current experimental measurements provide the following  upper limits~\cite{Hayasaka:2010np}:
\begin{align}
\mathcal{B}(\tau\to 3\mu)< 2.1 \times 10^{-8}\,, \\
\mathcal{B}(\tau\to 3e)< 2.7\times 10^{-8}\,.
\end{align}

\subsection{Lepton anomalous magnetic moment}

The scalar leptoquark couplings to charged leptons and quarks  give rise to an anomalous magnetic moment very similar to the contribution of the radiative decays $\ell^\beta\to\ell^\alpha \gamma$. The corresponding NP contributions to $a_\ell \equiv\frac{1}{2}\left(g-2\right)_\ell$ is given by~\cite{Choudhury:2001ad,Cheung:2001ip,Mandal:2019gff}
\begin{eqnarray}
\label{eq:g-2}
\Delta a_{\ell_\alpha}\, &=&\, \frac{-3}{16 \pi^2} 
\frac{m_\ell^2}{M^2}\,\sum_{j=u,c,t} \Bigg\{ \bigg( |g_3|^2 (V^*\tilde S_{QL})^{j\alpha} (V \tilde S_{QL}^*)^{j\alpha} +  |g_1|^2 (V^* S_{QL})^{j\alpha} (V S_{QL}^*)^{j\alpha}  \bigg)
\nn \\ 
&&\hskip 4cm\mbox{} \times 
\bigg[ -\frac{2}{3} F_{1}(x_j) + \frac{1}{3} F_{2}(x_j)\bigg] 
\nn \\ 
&&\hskip 1.2cm\mbox{}- m_{q_j} \sum_{j=u,c,t} \Re\,[ g_1 g_R^*\,
(V^*S_{QL})^{j\alpha}\, S_{UE}^{*j\alpha}] \, \bigg[ -\frac{2}{3} F_{3}(x_j) + \frac{1}{3}
F_4 (x_j)\bigg]\Bigg\}\,,
\end{eqnarray}
where the loop functions are defined in Eq.~\eqref{eq:loop_lgamma}.
It is apparent that the dominant contribution arises from the top quark in the loop in the presence of both left- and right-handed couplings.

The experimental measurement~\cite{Bennett:2006fi} and the theoretical prediction~\cite{Aoyama:2020ynm} currently show a discrepancy at a significance of 3.7$\sigma$:
\bea
\Delta a_\mu= a_\mu^{\rm exp} -a_\mu^{\rm SM} = (2.79 \pm 0.76) \times 10^{-9}\,.
\eea

\bibliographystyle{utphys}
\bibliography{references}

\end{document}